\documentclass[5p,times]{elsarticle}
\usepackage{algorithm}%
\usepackage{hyperref}
\usepackage{algpseudocode}%
\usepackage{amssymb}
\usepackage{graphicx}
\usepackage{pifont}
\usepackage{booktabs}          % For \toprule etc.
\usepackage{multirow}
\usepackage{adjustbox}
\usepackage{amsmath}
\usepackage{array} 
\usepackage{natbib}
\usepackage{caption}
\usepackage{amsmath,amssymb,amsfonts}%
\usepackage{url}
\usepackage{amsmath,amsthm}

\usepackage{amsmath}
\begin{document}

\begin{frontmatter}



\title{SORA-ATMAS: Adaptive Trust Management and Multi-LLM Aligned Governance for Future Smart Cities}


\author[1]{Usama Antuley}%[<options>]

%\cormark[1]


\ead{usama.antuley@nu.edu.pk}


%\credit{Writing – original draft, Conceptualization, Investigation, Methodology, Visualization, Formal analysis.}

\author[1]{Shahbaz Siddiqui}%[<options>]

\ead{shahbaz.siddiqui@nu.edu.pk}

%\credit{Writing – original draft, Conceptualization, Investigation, Methodology, visualization, Supervision.}

\author[1]{Sufian Hameed}%[]

% Footnote of the second author

% Email id of the second author
\ead{sufian.hameed@nu.edu.pk}


%\credit{Writing – review \& editing, Supervision, Conceptualization.}

\author[1]{Waqas Arif}%[<options>]

%\cormark[1]
\ead{waqas.arif.v@nu.edu.pk}
%\credit{Writing – original draft, Conceptualization, Investigation, Formal analysis.}

\author[2]{Subhan Shah}%[<options>]


\ead{shahsubhan708@gmail.com}

\author[3]{Syed Attique Shah}%[<options>]

%\cormark[1]


\ead{syedattique.shah@bcu.ac.uk}


%\credit{Writing – review \& editing, Supervision, Conceptualization, Funding acquisition}

\affiliation[1]{organization={Department of Computer Science, National University of Computer \& Emerging Sciences},
            addressline={St-4 Sector 17-D On National Highway}, 
            city={Karachi},
%          citysep={}, % Uncomment if no comma needed between city and postcode
            postcode={75160}, 
            state={},
            country={Pakistan}}

\affiliation[2]{organization={Balochistan University of Information Technology, Engineering and Management Sciences},
            addressline={Airport Road, Baleli}, 
            city={Quetta},
%          citysep={}, % Uncomment if no comma needed between city and postcode
            postcode={87300}, 
            %state={},
            country={Pakistan}}

\affiliation[3]{organization={Department of Computer Science, Birmingham City University},
            addressline={STEAMhouse, Belmont Row}, 
            city={Birmingham},
%          citysep={}, % Uncomment if no comma needed between city and postcode
            postcode={B4 7RQ},
          %  state={},
            country={United Kingdom}}

\cortext[1]{Corresponding author}


\begin{abstract}
The rapid evolution of smart cities has increased the reliance on intelligent interconnected services to optimize infrastructure, resources, and citizen well-being. Agentic AI has emerged as a key enabler by supporting autonomous decision-making and adaptive coordination, allowing urban systems to respond in real time to dynamic conditions. Its benefits are evident in areas such as transportation, where the integration of traffic data, weather forecasts, and safety sensors enables dynamic rerouting and a faster response to hazards. However, its deployment across heterogeneous smart city ecosystems raises critical governance, risk, and compliance (GRC) challenges, including accountability, data privacy, and regulatory alignment within decentralized infrastructures. Evaluation of \emph{SORA-ATMAS} with three domain agents (Weather, Traffic, and Safety) demonstrated that its governance policies, including a fallback mechanism for high-risk scenarios, effectively steer multiple LLMs (GPT, Grok, DeepSeek) towards domain-optimized, policy-aligned outputs, producing an average MAE reduction of $\approx 35\%$ across agents. Results showed stable weather monitoring, effective handling of high-risk traffic plateaus ($R \approx 0.85$), and adaptive trust regulation in Safety/Fire scenarios ($\tau_t = 0.65$). Runtime profiling of a 3-agent deployment confirmed scalability, with throughput between 13.8-17.2 requests per second, execution times below 72~ms, and governance delays under 100~ms; analytical projections suggest maintained performance at larger scales. Cross-domain rules ensured safe interoperability, with traffic rerouting permitted only under validated weather conditions. These findings validate \emph{SORA-ATMAS} as a regulation-aligned, context-aware, and verifiable governance framework that consolidates distributed agent outputs into accountable, real-time decisions, offering a resilient foundation for smart-city management.
\end{abstract}



%% Keywords

\end{frontmatter}



\section{Introduction}

The rapid development of smart cities has led to a growing reliance on interconnected and intelligent services that improve urban infrastructure, resource management, and citizen well-being. These services integrate technologies such as the Internet of Things (IoT), big data analytics, and artificial intelligence (AI) to create efficient urban ecosystems ~\cite{rathore2016urban, sarker2022smart}. Smart cities address challenges such as population density, energy consumption, and service delivery, fostering sustainable and responsive urban environments by enabling real-time data exchange and automated processes \cite{adeleke2025comprehensive, bittencourt2024survey, li2022big}. Recent advances emphasize the role of AI in optimizing these services, as seen in studies on sustainable urban planning where interconnected systems reduce carbon footprints, improve air quality, and improve overall quality of life through predictive modeling and resource allocation \cite{al2025smart}.

Agentic AI plays a key role in this evolution by providing autonomous decision-making and adaptive coordination, allowing systems to respond effectively to changing urban conditions. Unlike conventional AI, agentic systems can set goals, interact with environments, and collaborate independently, making them suitable for complex smart city operations \cite{tiwari2025agentic, murugesan2025rise}. This autonomy is essential for handling dynamic data from various sources, improving efficiency and scalability in urban settings \cite{Acharya2025}. For example, in the energy sector, agentic AI can play a vital role by managing consumption more intelligently, forecasting future demand, and improving the performance of renewable sources through the use of past usage trends and weather insights. Similarly, within transportation, it can improve routing efficiency, shorten delivery durations, and strengthen supply chain operations by continuously analyzing traffic flows, climatic conditions, and real-time vehicle data \cite{HOSSEINI2025}.

\begin{figure*}[t]
    \centering
    \includegraphics[width=0.58\linewidth]{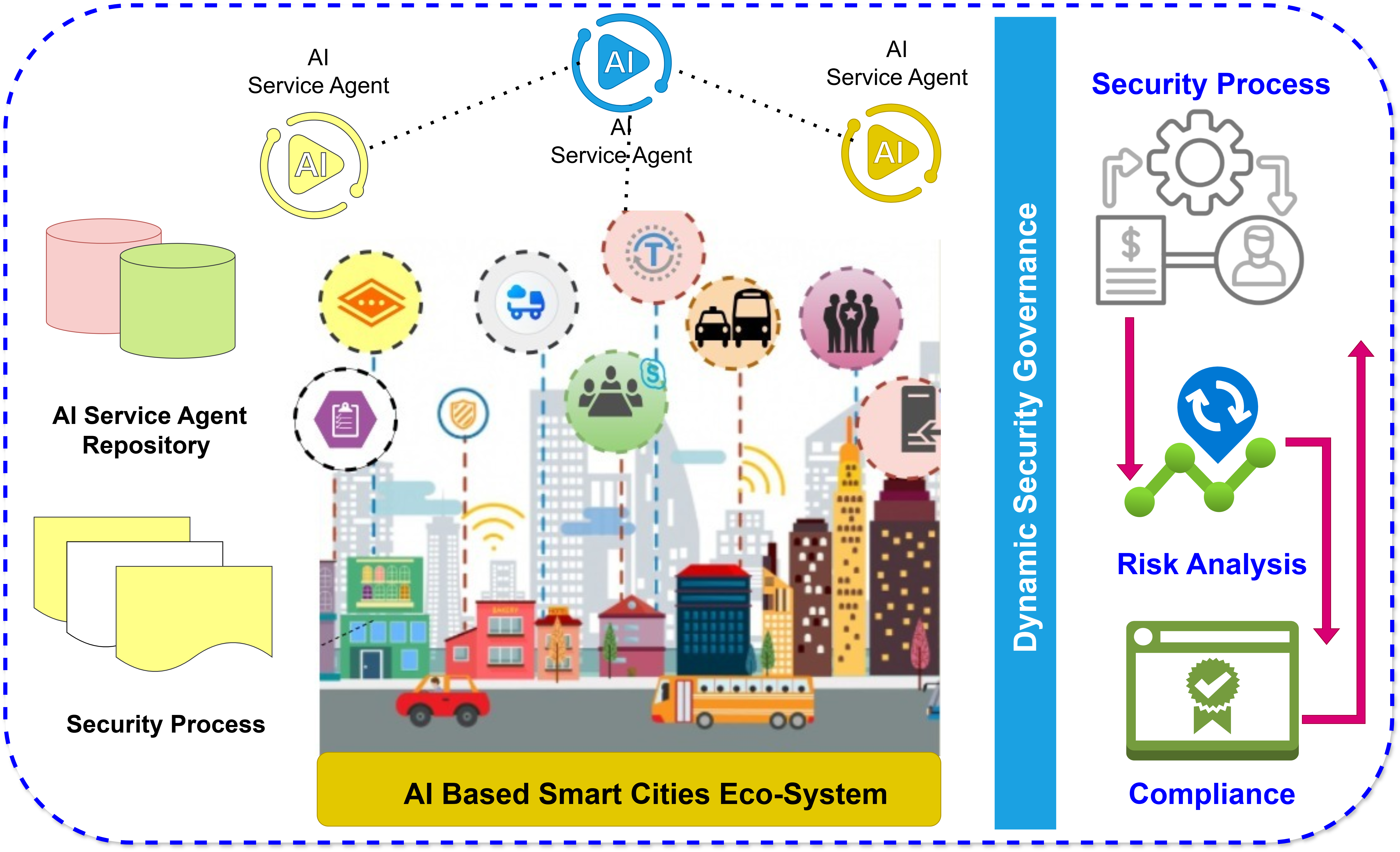}
    \caption{AI-based smart-city ecosystem illustrating agentic service agents and repositories supervised by a \emph{dynamic security governance} pipeline: security process → risk analysis → compliance, which coordinates cross-domain services via inter-agent links.}
    \label{fig:grc-diagram}
\end{figure*}

However, adopting agentic AI in heterogeneous smart city ecosystems brings notable governance, risk, and compliance (GRC) challenges. Governance requires defining clear accountability for autonomous actions, as unclear decision processes can lead to ethical issues and lack of transparency \cite{homsy2019multilevel, LARTEY2025}. Risks include algorithmic biases that may worsen social inequalities and vulnerabilities to cyberattacks in linked networks, potentially compromising system integrity and leading to data breaches \cite{homaei2024, Raj2023}. Compliance involves meeting regulations such as data privacy laws (e.g., GDPR) and AI ethics guidelines, which are often not designed for agentic systems' flexibility, requiring adaptive frameworks to ensure legal alignment amid evolving standards \cite{Singhal2024, Grace2025, smuha2025regulation}. These issues can erode trust and slow AI integration if not managed well, as evidenced by recent analyses of AI governance in urban contexts that highlight the need for robust oversight to prevent societal harm \cite{sanchez2025ethical, PRAHARAJ2025}.
Existing research often examines isolated aspects of agentic AI or GRC but lacks integrated frameworks that combine adaptive trust management with centralized and decentralized mechanisms for heterogeneous environments \cite{Karim2025,RIAD2021,rosmaninho2025}.

Blockchain has emerged as a critical enabler of governance, risk, and compliance (GRC) in smart cities due to its transparency, immutability, and decentralized control. Unlike centralized databases, it ensures that records of agentic AI decisions and data exchanges remain tamper-proof and auditable, which is vital for accountability and trust \cite{musik2025, islam2025trust}. Studies highlight its role in mitigating algorithmic opacity by providing verifiable decision logs for real-time auditing, improving explainability and reducing risks of bias or manipulation \cite{nassar2020blockchain, LIU2024}. Smart contracts further enable decentralized policy enforcement, supporting automated compliance checks and revocation across heterogeneous services without reliance on a single authority \cite{SIDDIQUI2024}. In terms of privacy, blockchain facilitates selective disclosure and secure multi-party transactions, aligning AI operations with regulations such as GDPR and ethics guidelines \cite{mustafa2025blockchain}. Overall, blockchain embeds GRC principles into the technological foundation of smart cities, strengthening resilience and citizen trust.

Addressing these challenges requires an integrated approach that can dynamically evaluate trust and risk across heterogeneous smart city services. Such a system should combine mechanisms for contextual assessments of agent decisions, incorporate governance agents to ensure coordinated operations, and employ decentralized technologies like blockchain to provide secure and transparent data exchange. By embedding accountability, adaptability, and resilience into the technological foundation of smart cities, future frameworks can better manage governance, risk, and compliance while enabling reliable and trustworthy agentic AI operations. Figure~\ref{fig:grc-diagram} provides a visual overview of the interplay between GRC principles and smart city domains.

\subsection*{Motivation}
The integration of Agentic AI into smart city infrastructures presents transformative opportunities for optimizing urban services such as transportation, energy, healthcare, and public safety \cite{LARTEY2025, HOSSEINI2025}. By autonomously coordinating data from IoT devices, cloud systems, and urban platforms, intelligent agents can enhance efficiency, adaptability, and resilience across domains \cite{tiwari2025agentic}. However, this autonomy also introduces vulnerabilities: biased or incomplete data may lead to misinformed decisions, errors can propagate rapidly across interconnected services, and accountability for agent actions often remains unclear in decentralized environments \cite{homaei2024, homsy2019multilevel}. Traditional governance approaches are insufficient to address these challenges \cite{SIDDIQUI2023}. Therefore, ensuring the trustworthiness, transparency, and compliance of Agentic AI requires adaptive trust management mechanisms, blockchain-enabled governance, and orchestrator agents (e.g., SORA) capable of dynamically validating outputs and regulating agent autonomy in real time \cite{islam2025trust, SIDDIQUI2024}. Based on this motivation and the reviewed literature, we formulate the following research questions:  

\begin{enumerate}
    \item RQ1: How can Agentic AI systems in smart cities be continuously monitored and validated to ensure accuracy, reliability, and fairness across heterogeneous domains?

    \item RQ2: In what ways can blockchain-based mechanisms establish decentralized governance, transparency, and accountability for multi-agent smart city ecosystems?

    \item RQ3: How can adaptive trust management and specialized orchestration agents dynamically regulate agent autonomy and coordination to strengthen resilience and compliance in smart city operations?
\end{enumerate}

\subsection*{Contribution}
The major contributions of this paper are outlined below, emphasizing both the novelty of the framework and its empirical validation against relevant baselines:

\begin{enumerate}
    \item We introduce \emph{SORA-ATMAS}, a multi-agent adaptive trust management framework that unifies decentralized agentic sensing and reasoning with centralized governance validation through SORA. This hybrid approach enables real-time, context-aware risk and trust evaluation at the edge while preserving accountability and compliance at the city level.

    \item The framework employs a dual-chain architecture, coupling Agentic Blockchains at the edge with the SORA Blockchain at the governance layer. Unlike prior blockchain-GRC systems focused on auditability, this setup ensures tamper-resistance, transparent compliance, and verifiable cross-domain coordination, preventing single points of failure and reinforcing accountability.

    \item To overcome the limitations of single-model systems, we design a multi-LLM evaluation mechanism integrating GPT, DeepSeek, and Grok as agent reasoning engines. Unlike online shielding \cite{konighofer2023} or static XACML policies \cite{XACML12019}, our ensemble combines MAE-based selection, governance thresholds, and iterative error feedback to enable adaptive convergence across domains, achieving up to \textbf{35\%} MAE reduction across agents.

    \item SORA-ATMAS is validated in a smart-city disaster management scenario with weather, traffic, and safety agents. The framework consistently converges towards the SORA baseline, demonstrating robustness under high-risk conditions (e.g., traffic plateaus with $R \approx 0.85$) and ensuring strong cross-agent interoperability, establishing its scalability and reliability for heterogeneous smart-city ecosystems.

\end{enumerate}

The remainder of this paper is organized as follows: Section 2 analyzes the security challenges associated with governance, risk, and compliance (GRC) for AI systems in heterogeneous urban environments. Section 3 provides a comprehensive literature review covering agentic AI, adaptive trust, and blockchain-enabled GRC models. Section 4 outlines the overall system design of the proposed SORA-ATMAS framework, followed in Section 5 by a detailed use case on decentralized, trusted GRC for smart-city disaster management. Section 6 explains the workflow execution across agents and governance layers, demonstrating how adaptive trust and compliance are operationalized. Section 7 presents results and discussion, highlighting empirical findings and convergence behaviors, while Section 8 concludes the paper and outlines future directions.

\section{Adaptive Governance, Risk \& Compliance for AI}
Building on the introduction's exploration of agentic AI's potential in smart cities, this section examines Governance, Risk, and Compliance (GRC) for AI systems. GRC frameworks are vital for ensuring responsible AI use and compliance with urban standards. We review centralized GRC models, their security issues, and necessary improvements, including decentralization, accountability, trust management, and agents like SORA for adaptive governance. Centralized GRC, the most common model, aligns AI-enabled services with urban policies through regulatory agencies or municipal governments \cite{smuha2025regulation}. These models govern sectors like energy, transportation, and surveillance under coordinated compliance and supervisory controls. Traditional implementations rely on institutional monitoring, dashboards, audits, and penalties to ensure AI systems align with urban policies \cite{odedina2023redefining}. While offering oversight, consistent policy application, and regulatory alignment across heterogeneous systems \cite{sadeghi2024interoperability}, centralized governance also enhances risk handling and harmonizes AI service deployment in urban planning.

\subsection{Security Challenges in Centralized GRC for AI}
Although centralized GRC offers benefits such as consistency and standardization, it also introduces several critical security challenges when applied to complex smart-city ecosystems:  

\begin{enumerate}
    \item The central governance hub represents a vulnerable bottleneck, where disruption or compromise, whether caused by technical malfunction or deliberate cyberattack i.e, including denial-of-service (DoS) attacks, can trigger cascading failures across multiple interconnected urban services \cite{Raj2023, Alvin2020}.

    \item Scalability becomes a major concern as IoT devices and agentic AI systems continuously generate massive volumes of data. Centralized risk evaluation mechanisms are often unable to scale or adapt in real time, leaving urban infrastructures exposed to unpredictable and evolving risks \cite{SIDDIQUI2024}.  

    \item Adversaries may engage in \emph{sensor spoofing}, i.e, falsified temperature, humidity, or traffic counts to distort risk assessments, or exploit \emph{LLM prompt injection} to manipulate agent reasoning outputs. Both actions compromise decision quality and undermine trust in AI-driven governance \cite{DEVITO2025, adewusi2024conceptual}.  
    
    \item Privacy and data security risks are intensified by storing sensitive citizen and operational records in a single repository. This centralization increases the likelihood of large-scale breaches, unauthorized access, and regulatory violations, raising concerns about both individual rights and institutional liability \cite{smuha2025regulation}.  
\end{enumerate}

These challenges underscore why recent critiques argue that centralized GRC cannot meet the flexibility, resilience, and accountability requirements of heterogeneous multi-agent smart-city ecosystems \cite{SIDDIQUI2023}.

\subsection{Security Requirements for Adaptive GRC}
Overcoming the above limitations requires a shift from rigid, centralized control toward more adaptive governance that integrates decentralization and distributed resilience, while still preserving central oversight where necessary. The following requirements directly respond to the identified challenges:  

\begin{enumerate}
    %\item Decentralization of oversight is needed to mitigate the risks of a vulnerable bottleneck. By distributing compliance and governance functions across multiple nodes, reliance on a single point of failure is reduced and systemic resilience is enhanced. Intermediary orchestration agents, such as SORA, further balance centralized oversight with distributed autonomy, ensuring resilience against targeted attacks and enabling rapid recovery from localized disruptions \cite{antuley2025securing, SIDDIQUI2023, islam2025trust}.  

     \item Decentralization of oversight reduces reliance on a single governance hub and increases systemic resilience against DoS attacks. By distributing compliance and governance across multiple nodes, intermediary orchestration agents such as SORA ensure continuity even under targeted disruption \cite{antuley2025securing, SIDDIQUI2023, islam2025trust}

    \item Governance mechanisms must process high-volume, rapidly evolving IoT and AI data while also validating semantic reasoning outputs. In SORA-ATMAS, MAE-based validation and iterative feedback loops counter risks such as LLM prompt injection, ensuring that manipulated outputs cannot bypass policy baselines \cite{dazzi2025internet}.

    \item Flexible and context-aware policy frameworks are required to overcome rigid enforcement. These frameworks permit localized adaptation while maintaining global standards, ensuring responsiveness in heterogeneous and dynamic environments \cite{golpayegani2024adaptation}.  

%Flexible and context-aware frameworks enable localized adaptation while maintaining global standards. By enforcing threshold checks $(\epsilon_R, \epsilon_T)$ and corroborating signals across agents, SORA-ATMAS detects and contains sensor spoofing attempts, ensuring reliable data integrity.  
    
    \item Stronger privacy and accountability safeguards should be implemented through secure audit trails, distributed data storage, access control measures, and clear responsibility mapping. This reduces the risks of large-scale breaches, strengthens compliance with regulatory standards, and enhances transparency in AI-driven decision-making to build public trust \cite{SIDDIQUI2024, Ayub2024}.  
\end{enumerate}

Taken together, these requirements form the basis of an adaptive and hybrid GRC architecture. Such a model balances centralized accountability with decentralized resilience, enabling secure, transparent, and context-aware governance across smart-city infrastructures, thereby enhancing both public trust and operational robustness.

\section{Literature Review}

Smart cities are increasingly framed as cyber-physical ecosystems in which sensing, connectivity, and AI coordinate resources and services at urban scale \cite{antuley2025securing}. Empirical studies demonstrate how deep and reinforcement-learning pipelines are reshaping urban operations: city-scale traffic signal optimization with multi-agent RL controllers compresses travel times and reduces congestion \cite{Tong2020,LI2021,cai2024traffic}, while multimodal prediction models fusing air quality, meteorological, and traffic data advance public-health and climate agendas \cite{hameed2023}. Autonomous multi-agent systems, such as decentralized rendezvous planning for robots, enhance urban functionality in smart cities, supporting operations like search-and-rescue, surveillance, and assembly \cite{Ozsoyeller2024}. Collectively, these results show that AI-enhanced programs lower lifecycle costs through predictive maintenance, streamline mobility by coordinating heterogeneous flows, and support sustainability targets via precise forecasting.

Building on such assistive gains, research is now pivoting toward agentic AI systems that reason, plan, and act with minimal supervision, often organized as multi-agent systems (MAS). In operational domains, MARL coordinates heterogeneous services in real time: agents negotiate priorities across distributed intersections, allocate resources, and trigger cross-system actions under uncertainty \cite{LI2021,cai2024traffic}. Architecturally, edge-cloud designs push computation closer to sensors and vehicles for low-latency safety-critical tasks, while cloud layers synchronize global state and policy. Standardized oneM2M-MEC interworking and Modbus IoT gateways exemplify interoperability for legacy and modern services \cite{LEE2025int, ELAMANOV2024mod}. These patterns improve responsiveness but complicate enforcement and explainability, motivating structured GRC frameworks. Agentic models also begin to probe counterfactual scenarios for proactive planning \cite{cai2024traffic}, highlighting adaptability yet raising governance challenges.

As cities move from automation to autonomy, governance, risk, and compliance (GRC) emerge as critical enablers of trustworthy operations. Smart-contract controls with tamper-evident blockchain logging support authentication, authorization, and forensic accountability across services \cite{SIDDIQUI2023}, though scalability and policy rigidity remain concerns in large ecosystems. Adaptive governance extends these mechanisms with on-chain policy compilation and runtime compliance metrics \cite{SIDDIQUI2024}, but incurs significant overhead in multi-domain settings. Decentralized trust frameworks that integrate blockchain with AI-driven techniques strengthen decisions in safety-critical environments \cite{islam2025trust}, yet introduce latency, energy costs, and lack regulatory pathways. Online shielding provides embedded compliance by blocking unsafe actions during learning and execution \cite{Koenighofer2023}, though adaptability across complex MAS is limited. Complementary centralized approaches include adaptive XACML for runtime-aware IoT access control \cite{RIAD2021}, though growing policy complexity challenges robustness. Similarly, IRM/GRC suites combined with blockchain and AI analytics enhance resilience \cite{Ali2025, Ayub2024}, but high costs, vendor lock-in, and untested scalability hinder adoption. At the urban scale, Integrated City Command and Control Centers (ICCCs) demonstrate both the promise and risks of centralized governance, improving oversight but also exposing vulnerabilities through corporate dominance, uneven implementation, and limited public accountability \cite{PRAHARAJ2025}.

Interoperability magnifies these pressures. Smart-city infrastructures braid legacy assets, vendor-specific protocols, and fragmented schemas, creating bottlenecks for cross-domain coordination. Standards-based edge-cloud interworking \cite{LEE2025int,ELAMANOV2024mod} and digital twins that integrate real-time mobility, environment, and utility data provide shared substrates for decision-making \cite{sohail2025udt}. However, centralized hubs that unify standards can also become throughput bottlenecks, slowing response and reducing locality. This motivates adaptive governance that delegates bounded authority to districts, utilities, and agents, while preserving alignment with citywide policy objectives.

Despite notable progress, two enduring gaps continue to constrain the maturation of agentic smart-city systems. First, the linkage between explainability and governance remains underdeveloped: existing mechanisms fall short of producing verifiable evidence that autonomous decisions conform to legal, ethical, and operational norms. Action-space enforcement through online shielding represents a promising advance in runtime compliance, yet its application is still narrow and not generalized across multi-agent platforms \cite{Koenighofer2023}. Second, achieving equity and resilience at the city scale requires institutional innovation alongside technical integration. Recent studies highlight the need for standardized edge–cloud orchestration to ensure reliable coordination of mission-critical services \cite{rosmaninho2025}, transparent digital-twin pipelines to maintain observability over dynamic urban processes \cite{sohail2025udt}, and trust frameworks with measurable guarantees for both security and benefit sharing \cite{islam2025trust}. Collectively, these insights converge toward a hybrid governance trajectory: maintaining centralized GRC for standardization and high-risk veto authority, while delegating execution to decentralized, context-aware mechanisms that can address the scale, heterogeneity, and resilience demands of agentic smart-city ecosystems. Table~\ref{tab:litreview} summarizes the coverage of these studies across key smart-city domains, agentic AI integration, interoperability, and governance models.

\begin{table*}[h]
    \caption{Coverage of reviewed studies across smart-city domains, AI and Agentic AI integration, compliance, and governance dimensions. A check mark ($\checkmark$) indicates substantive coverage; $\times$ indicates not a primary focus.}
    \label{tab:litreview}
    \centering
    \begin{adjustbox}{width=\textwidth,keepaspectratio}
    \begin{tabular}{|
        p{1.6cm}|
        >{\centering\arraybackslash}p{3.3cm}|
        >{\centering\arraybackslash}p{1.7cm}|
        >{\centering\arraybackslash}p{1.7cm}|
        >{\centering\arraybackslash}p{2.1cm}|
        >{\centering\arraybackslash}p{2.3cm}|
        >{\centering\arraybackslash}p{2.3cm}|}
        \hline
        \textbf{Study} &
        \textbf{Smart City Domain} &
        \textbf{Agentic AI Integration} &
        \textbf{Trust \& Risk Management.} &
        \textbf{Security \& Operational Compliance} &
        \textbf{Heterogeneous Services Interoperability} &
        \textbf{Governance\ \& Compliance} \\ \hline
        
        \cite{WANG2021} & Traffic Control & $\checkmark$ & $\times$ & $\times$ & $\times$ & $\times$  \\ \hline
        \cite{cai2024traffic} & Traffic Signal Optimization & $\checkmark$ & $\times$ & $\times$ & $\times$ & $\checkmark$  \\ \hline
        \cite{hameed2023} & Air Quality + Traffic Analytics & $\checkmark$ & $\checkmark$ & $\times$ & $\times$ & $\times$  \\ \hline
        \cite{ELAMANOV2024mod} & IoT / Industrial Services & $\checkmark$ & $\times$ & $\checkmark$ & $\checkmark$ & $\times$  \\ \hline
        \cite{LEE2025int} & IoT + Edge Platforms & $\checkmark$ & $\times$ & $\checkmark$ & $\checkmark$ & $\checkmark$  \\ \hline
       \cite{SIDDIQUI2023} & Municipal Cross-domain Services & $\times$ & $\checkmark$ & $\checkmark$ & $\checkmark$ & $\checkmark$ \\ \hline
        \cite{SIDDIQUI2024} & Security Governance & $\times$ & $\checkmark$ & $\checkmark$ & $\checkmark$ & $\times$ \\ \hline
       \cite{islam2025trust} & Blockchain Trust for Smart Cities & $\times$ & $\checkmark$ & $\checkmark$ & $\times$ & $\times$ \\ \hline
        \cite{Ayub2024} & Smart City IoT Security & $\times$ & $\checkmark$ & $\checkmark$ & $\times$ & $\checkmark$  \\ \hline
        \cite{PRAHARAJ2025} & Command \& Control (ICCC) & $\times$ & $\times$ & $\checkmark$ & $\times$ & $\checkmark$ \\ \hline
        \cite{sohail2025udt} & Urban Digital Twins & $\checkmark$ & $\times$ & $\times$ & $\checkmark$ & $\times$  \\ \hline
        \cite{rosmaninho2025} & Edge-Cloud Orchestration & $\checkmark$ & $\times$ & $\checkmark$ & $\checkmark$ & $\checkmark$  \\ \hline
         Our Framework & Multiple City Domains & $\checkmark$ & $\checkmark$ & $\checkmark$ & $\checkmark$ & $\checkmark$ \\ \hline

    \end{tabular}
    \end{adjustbox}
\end{table*}

\section{System Overview}
\label{sec:system_overview}

The proposed framework adopts a layered Software Defined Internet of Things (SDIoT) architecture that establishes the structural foundation for adaptive, agent-driven governance in smart-city ecosystems. Within this design, SDIoT is not positioned as the core novelty but serves as an enabling abstraction that provides modularity, programmability, and scalable orchestration across heterogeneous IoT infrastructures. Its layered organization maintains a clear separation between perception, control, and application functions while facilitating integrated trust evaluation, decentralized policy enforcement, and city-wide accountability.
\begin{enumerate}
    \item At the top, the \textit{Application Layer} incorporates both the SORA Governance Layer and the Agentic Layer. The SORA Governance Layer operates as the central authority for city-wide oversight, embedding engines for security policy dissemination, cross-domain coordination, adaptive trust and GRC enforcement, and blockchain-based governance repositories. In parallel, the Agentic Layer enables domain-specific autonomy through decentralized agents such as those for weather, traffic, and safety that integrate compliance mechanisms, contextual reasoning via LLMs, and local blockchain anchoring. Together, these modules balance local adaptability with global consistency, ensuring that autonomous services remain regulation-aligned, verifiable, and context-aware.

      \item Beneath the application plane lies the \textit{control layer}, which manages communication, heterogeneity, and scalability through SDN-inspired mechanisms such as SDN-WISE controllers. Multiple controllers can be deployed with failover to maintain reliability, with their number determined by load and redundancy needs. If the primary controller fails, backups are automatically activated to ensure uninterrupted service. By separating control and data planes, this layer enables programmability, topology discovery, and flow management across diverse IoT domains. Operating over IEEE 802.15.4, packets are handled via the WISE Flow Table for interoperability. It also manages cryptographic keys and sessions, generating ECC key pairs (128/192/256 bits) and distributing session keys using Elliptic Curve Diffie–Hellman (ECDH). ECC curves (i.e. secp256r1, secp384r1) and key rotation intervals (i.e. every 20 days or after a set number of transactions) are defined per security policy. These cryptographic measures ensure authenticated, encrypted communication and secure enforcement of high-level policies.

    \item At the foundation lies the \textit{perception layer}, composed of distributed IoT devices such as environmental sensors (temperature, humidity, wind, rainfall), traffic monitoring systems, and safety detectors or cameras for fire/smoke detection. These devices continuously capture real-time data, which is pre-processed and logged into structured repositories. The collected observations provide the basis for trust evaluation, risk estimation, and governance enforcement. By delivering continuous situational data (e.g., weather anomalies, traffic congestion, visible smoke), this layer supports disaster management, mobility optimization, and public safety. Each IoT node is cryptographically identified through the control layer, ensuring data integrity and secure participation.  
\end{enumerate}

\begin{figure*}[h]
    \centering
    \includegraphics[scale=0.4]{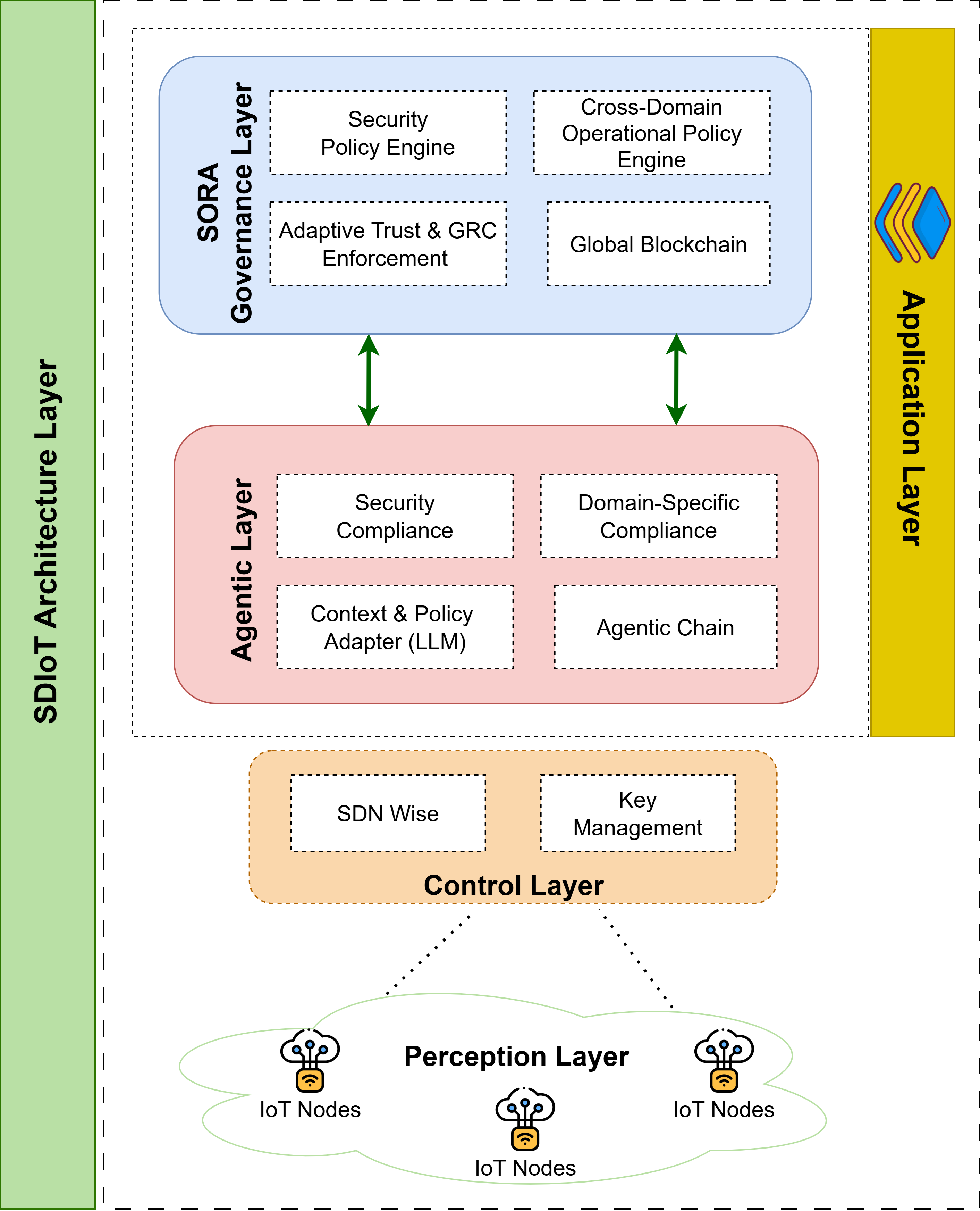}
    \caption{Layered SDIoT-based governance framework integrating the SORA Governance Layer and Agentic Layer within the Application Layer.}
    \label{fig:sdioT_framework}
\end{figure*}

Together, these components form the SDIoT Architecture Layer, the structural backbone of the system. As shown in Fig.~\ref{fig:sdioT_framework}, the layered design enables modular scalability and seamless integration of heterogeneous IoT services, while SORA governance and agentic intelligence modules enforce adaptive trust, risk-aware compliance, and resilient cross-domain collaboration. This organization harmonizes local autonomy with global oversight, ensuring secure, accountable, and adaptive governance in dynamic smart-city environments.

\subsection{Proposed Architecture}
\label{proposed_architecture}

\begin{figure}[h]
    \centering
    \includegraphics[scale=0.4]{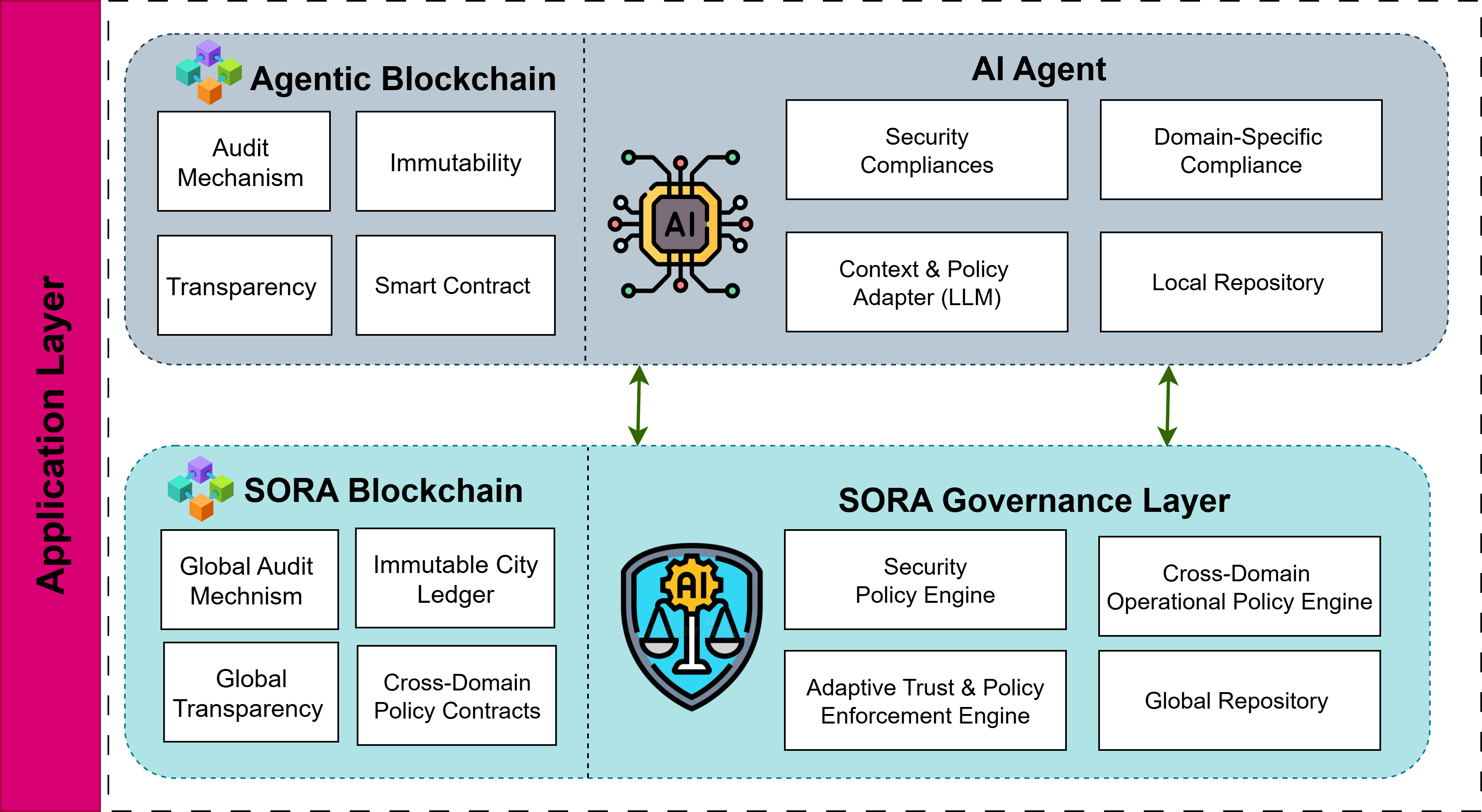}
    \caption{Proposed governance framework integrating decentralized AI agents, Agentic Blockchain, and the SORA Governance Layer over the SORA Blockchain, enabling adaptive trust and cross-domain policy enforcement.}
    \label{fig:framework_overall}
\end{figure}

The proposed architecture leverages a dual-chain governance model where decentralized AI agents operate autonomously at the domain level while remaining anchored to a global oversight framework. Figure~\ref{fig:framework_overall} illustrates the overall design, comprising domain-specific AI agents, the SORA Governance Layer, and the dual-chain blockchain substrate. This design ensures that local intelligence, contextual adaptability, and regulatory compliance are synchronized with city-wide accountability and interoperability.

\subsection{Agentic Layer}
AI agents form the operational core of the architecture lies in the Application Layer, representing heterogeneous smart-city services such as mobility, healthcare, energy, environment, or public safety. Each agent is designed to operate autonomously within its respective domain, yet remains accountable to both local regulatory constraints and overarching city governance rules. Unlike static service modules, these agents embed compliance intelligence and adaptive reasoning, enabling them to respond dynamically to contextual changes, emerging risks, and updated policy directives.  
 
The AI agent integrates four tightly coupled components:  
\begin{itemize}
    \item \textbf{Security Compliance:} Enforces core protocols such as authentication, authorization, encryption, and integrity validation. It ensures only legitimate services register, authenticated entities exchange data, and communications remain cryptographically secured. To mitigate emerging risks, it supports optimal ECC selection, dynamic re-keying, and SHA-256 integrity checks safeguarding against impersonation, tampering, and unauthorized access.

    \item \textbf{Domain-Specific Compliance:} Embeds sector-specific regulatory logic to meet operational and legal mandates. For instance, healthcare agents enforce GDPR confidentiality, traffic agents uphold real-time safety thresholds, and energy agents apply efficiency and sustainability benchmarks. By integrating these rules into compliance, agents ensure functional optimization while satisfying governance and legal obligations.  

    \item \textbf{Context and Policy Adapter (LLM):} Uses large language models to interpret contextual signals from IoT streams, logs, or requests. It semantically assesses alignment with baseline policies and, when anomalies arise (e.g., congestion, abnormal vitals, hazards), updates or recommends enforcement actions such as restricting access, reallocating resources, raising encryption tiers, or issuing alerts. This LLM-driven intelligence transforms agents into adaptive compliance actors.  

    \item \textbf{Local Repository:} The Local Repository is a secure data storage system that stores validated policies, compliance actions, and risk-trust scores. It uses a traditional DBMS for non-sensitive data. Retention policies ensure that sensitive data, including PII from traffic and CCTV sources, is anonymized or deleted after a set period (e.g., 30 days). PII minimization techniques like data aggregation, face masking, and encryption ensure GDPR compliance, while role-based access control (RBAC) limits access to authorized personnel.  
\end{itemize}

\subsection{SORA Governance Layer}
While AI agents provide localized autonomy and adaptive compliance, the SORA Governance Layer retains ultimate authority over their operation. It functions as the ruling governance entity, validating, approving, or overruling agent decisions to ensure consistency with city-wide objectives, systemic risk thresholds, and applicable regulatory mandates. Agents may autonomously generate proposals (e.g., escalation of encryption, restricting access, or adapting domain-specific rules), but these proposals are treated as \emph{recommendations} until they are formally reviewed and ratified by SORA. This hierarchical relationship ensures that responsiveness and local autonomy never compromise overall accountability, trust, or compliance.  

The SORA Governance Layer comprises four integrated engines, each mapped to a supervisory function:  

\begin{itemize}
    \item \textbf{Security Policy Engine:} Defines baseline directives on cryptographic mechanisms, authentication factors, and access control. All agent-level checks (e.g., ECC tier selection, ACLs) are validated against this global baseline, ensuring uniform security across the ecosystem.  

    \item \textbf{Cross-Domain Operational Policy Engine:} Enforces sectoral compliance before inter-operation. Agents must first satisfy local rules such as GDPR confidentiality in healthcare, mobility safety in traffic, or sustainability in energy. Once verified, the engine governs interoperability by applying high-level directives for example, regulating healthcare, mobility data exchanges, validating traffic emergency coordination, or enforcing energy environment retention rules. Sequencing local then cross-domain checks guarantees secure, regulation-compliant, city-wide interactions.  

    \item \textbf{Adaptive Trust and Risk Enforcement Engine:} Monitors contextual trust and aggregated risk across agents. Even if agents classify themselves as trustworthy, SORA can overrule when thresholds are breached, imposing controls such as scope restriction, suspension, or human oversight. This engine acts as the \emph{supreme risk governor}, ensuring trust is enforced globally, not just declared locally.  

    \item \textbf{Global Repository:} Serves as canonical governance memory, storing validated policies, registration records, risk--trust trajectories, and enforcement decisions. Anchored to the SORA Blockchain, it provides immutable, auditable, and city-wide accessible records, while synchronizing historical data back to agents for adaptation and learning.  
\end{itemize}

Together, these engines elevate SORA from static oversight to an \textit{active and adaptive ruling authority}. Local autonomy enables rapid responses, while systemic validation ensures transparency, accountability, and compliance. Admission and operational validation proceed as outlined in Algorithm~\ref{alg:policy_validation}.

\begin{algorithm}[h]
\caption{Policy Validation and Enforcement by SORA}
\label{alg:policy_validation}

\textbf{Input:} Agent proposal $P_i=\langle S_i,\ \text{action},\ R_i,\ T_i\rangle$, local policy $\pi_i$, global thresholds $(\theta_R,\theta_T)$, optional partner $S_j$ \\
\textbf{Output:} Decision $\in \{\texttt{approve},\texttt{restrict},\texttt{deny}\}$ anchored on SORA-Chain

\begin{algorithmic}[1]
\State \textbf{Security check (Security Policy Engine):}
Verify identity, authentication, authorization, and access control for $S_i$. If any fail $\rightarrow$ \texttt{deny}.
\State \textbf{Domain-specific compliance:}
Confirm that $S_i$ satisfies sectoral rules (e.g., GDPR confidentiality/minimization, mobility safety thresholds, energy sustainability/retention). If non-compliant $\rightarrow$ \texttt{deny}.
\State \textbf{Trust/risk gate:}
\If{$T_i < \theta_T$}
    \State decision $\gets \texttt{deny}$ \Comment{trust below threshold}
\ElsIf{$R_i > \theta_R$}
    \State decision $\gets \texttt{restrict}$ \Comment{limit scope or require monitoring/human review}
\Else
    \State decision $\gets \texttt{approve}$
\EndIf
\State \textbf{Cross-domain clause (Cross-Domain Operational Policy Engine):}
If proposal involves $S_j$, first ensure both agents pass their own domain-specific checks; then enforce cross-domain constraints (scope, data minimization, safety, retention). If any violation $\rightarrow$ decision $\gets \texttt{deny}$.
\State \textbf{Anchoring:}
Record $\{S_i,S_j,\text{action},R_i,T_i,\text{decision}\}$ in \texttt{GovDecisions} and append an immutable entry to the SORA Blockchain.
\State \Return decision to $S_i$ (and $S_j$ if applicable).
\end{algorithmic}
\end{algorithm}

\subsection{Adaptive Trust and GRC Enforcement} \label{metrics}
To operationalize governance across heterogeneous smart-city services, we formalize how each \emph{agent} quantifies (i) environmental risk, (ii) service reliability, (iii) contextual trust, and (iv) adaptive (overall) trust and risk, and how the governance node (SORA) enforces city-level policies with measurable tolerances. The constructs are domain-agnostic and apply to diverse smart-city services such as weather, mobility, utilities, health, and safety. The framework provides compact, auditable hooks from per-agent assessments to global orchestration and review.

\paragraph{Definition 1 (Environmental Risk): }
Environmental risk quantifies the degree to which real-time sensor or service observations deviate from normative operating conditions in a smart-city ecosystem. It applies uniformly across heterogeneous domains by adapting to three generic data modalities: continuous signals, capacity/volume conditions, and discrete hazard events. Formally:
\begin{equation}
\small
R_{\text{Env}}^{i}(t)=
\begin{cases}
\frac{1}{n}\sum_{k=1}^{n} I\!\big(|x_k-\mu_k|>\theta_k\big), & \text{continuous signals}, \\[3pt]
I \!\big(\text{Load}(t)>\theta_{\text{cap}}\big), & \text{capacity/volume conditions}, \\[3pt]
I \!\big(\text{HazardEvents}(t)\ge 1\big), & \text{discrete hazard events}.
\end{cases}
\end{equation}

\noindent In practice, environmental and climate risk may be triggered by abnormal temperature, humidity, wind, precipitation, or pollution levels; mobility and transport by road congestion, public transit occupancy exceeding 90\%, or saturated parking; utilities and infrastructure by electricity demand beyond 85\% of grid capacity, water consumption nearing allocation limits, or network bandwidth saturation; and public safety or health by verified fire/smoke alarms, biohazard events, or epidemic outbreak alerts.

\paragraph{Definition 2 (History–Reputation Trust; HRT):}
This measure reflects how much an agent can be trusted based on its past performance and peer consensus. It balances historical outcomes with external endorsements, smoothed over time:
\begin{equation}
\small
T_{\text{HRT}}^{i}(t)=
\begin{cases}
\alpha, & t=t_0,\\[4pt]
\delta\,T_{\text{HRT}}^{i}(t{-}\Delta T) +(1{-}\delta)\Big(\alpha\,s(t)+\beta\,T_{\text{Rept}}^{i}(t)\Big), & \text{otherwise},
\end{cases}
\end{equation}

where $\alpha=0.5$, $\beta=0.5$, $s(t)\in\{0,1\}$ is success/failure, and $\delta=0.85$ is the forgetting factor.

Peer reputation trust derives from model alignment within the ecosystem:
\begin{equation}
T_{\text{Rept}}^{i}(t)=
\frac{\sum_{j=1}^{M} c_j\, I(f_j=f_{\text{best}})}
{\sum_{j=1}^{M} c_j},\quad T_{\text{Rept}}^{i}(t_0)=0.5.
\end{equation}

\paragraph{Definition 3 (Service Risk):}
Service risk is defined as the complement of historical trust. It estimates the likelihood of a service failing in the near future based on its previous track record:
\begin{equation}
R_{\text{Service}}^{i}(t)=1-T_{\text{HRT}}^{i}(t{-}\Delta T),
\qquad R_{\text{Service}}^{i}(t_0)=0.5.
\end{equation}

\paragraph{Definition 4 (Overall Agent Risk):}
Each agent’s risk combines environment-driven anomalies and service reliability. A tunable parameter $\lambda_i$ determines the emphasis on each:
\begin{equation}
R^{i}(t)=\lambda_i\,R_{\text{Env}}^{i}(t)+(1{-}\lambda_i)\,R_{\text{Service}}^{i}(t),
\end{equation}
where $\lambda_i\in[0,1]$. Typical settings: environment $0.6$, mobility/utilities $0.7$, safety/health $0.8$.

\paragraph{Definition 5 (Contextual Trust):}
Contextual trust modifies a baseline confidence level based on operational factors such as data freshness, integrity, and compliance. It ensures agents degrade gracefully as conditions worsen:
\begin{equation}
T_{\text{Ctx}}^{i}(t)=
\min\!\Bigg(T_{\text{base}}
\prod_{k=1}^{n_i}(M_{i,k}(t))^{w_{i,k}},1.0\Bigg),
\end{equation}
where $T_{\text{base}}=0.7$, and modifiers $M_{i,k}$ capture dynamic context (sensor health, message integrity, threshold proximity).

\paragraph{Definition 6 (Overall Trust):}
Overall trust fuses historical and contextual components with adaptive weighting. As agent risk increases, contextual trust is emphasized more:
\begin{equation}
T_{\text{Overall}}^{i}(t)=
w_{\text{HRT}}(t)\,T_{\text{HRT}}^{i}(t)+w_{C}(t)\,T_{\text{Ctx}}^{i}(t),
\end{equation}
where
\[
w_{\text{HRT}}(t)=0.5-0.2\,R^{i}(t),\quad
w_{C}(t)=0.5+0.2\,R^{i}(t).
\]

\paragraph{Definition 7 (Ecosystem Metrics).}
At the city scale, ecosystem metrics aggregate trust and risk across all active agents. This provides SORA with system-level visibility:
\begin{equation}
\begin{aligned}
T_{\text{Ecosystem}}(t) &= \tfrac{1}{|\mathcal{A}(t)|}\sum_{i\in\mathcal{A}(t)} T_{\text{Overall}}^{i}(t), \\[4pt]
R_{\text{Ecosystem}}(t) &= \max_{i\in\mathcal{A}(t)} R^{i}(t).
\end{aligned}
\end{equation}

\subsection{Agentic and SORA Blockchains}
The governance framework adopts a dual-chain paradigm that couples localized assurance with city-wide oversight. At the domain level, each AI agent anchors decisions to an \textit{Agentic Blockchain}, a decentralized compliance ledger that guarantees immutability, enables fine-grained audits, and enforces sector-specific rules through smart contracts. At the global level, the \textit{SORA Blockchain} aggregates and harmonizes evidence from all domain chains, provides cross-domain audit trails, enforces shared policy contracts, and operationalizes governance via adaptive trust updates, escalations, and emergency procedures. Together, the Agentic and SORA chains form a layered trust backbone: Agentic Chains provide decentralized, context-sensitive compliance, while the SORA Chain consolidates these assurances into a globally auditable, risk-aware governance fabric, avoiding single points of failure and keeping local adaptability synchronized with city-wide objectives. 

The end-to-end enforcement flow that unifies per-agent execution and global validation is specified in Algorithm~\ref{alg:unified_enforcement} (with metric computations referencing Definitions~1–7). %and is visually summarized in Figure~\ref{fig:algo agentic-sora-flow}.

\begin{algorithm}[h]
\caption{Policy Enforcement and Logging via Dual-Chain Governance}
\label{alg:unified_enforcement}

\textbf{Input:} Agent observation for $S_i$ (\texttt{AgentLogs}); key $K_b$; thresholds $(\theta_R,\theta_T)$; domain baselines $\tau_T(i)$; tolerances $(\epsilon_R,\epsilon_T,\epsilon_{\text{tie}})$; stability $h$; cooldown $\Delta t$; optional partner $S_j$.  
Metrics computed per Definitions (Defs)~1--7 in Subsection~\ref{metrics}; policy conditions from Table~\ref{tab:sora-policy-matrix}.

\textbf{Output:} Decision $\in \{\texttt{approve},\texttt{restrict},\texttt{deny}\}$ anchored on Agent-Chain and SORA-Chain.

\begin{algorithmic}[1]
\State \textbf{Agent-side (W1--F1):} Authenticate $K_b$; preprocess input; compute $R_{\text{Env}}$, $T_{\text{HRT}}$, $R_{\text{Service}}$, $R^i$, $T_{\text{Ctx}}$, and $T_{\text{Overall}}$ (Defs.~1--6).  
Append results to \texttt{AgentLogs} and anchor record on \emph{Agent-Chain}. Forward packet $P_i=\{S_i,t,R^i,T_{\text{Overall}}^i\}$ to SORA.

\State \textbf{SORA ingress (S1):} Validate and admit only if packet is consistent within $(\epsilon_R,\epsilon_T)$ and $T_{\text{Overall}}^i \ge \tau_T(i)$, following gate criteria in Table~\ref{tab:sora-policy-matrix}.

\State \textbf{Selection \& feedback (S2--S3):} If multiple variants exist, select highest trust (tie $\rightarrow$ closest risk). Issue error-directed feedback ($\Delta R,\Delta T$) to non-selected candidates, per policy rules.

\State \textbf{Decision (policy thresholds):}  
\quad If $T_i < \theta_T \rightarrow$ \texttt{deny};  
\quad Else if $T_i < 0.7$ or $R_i > \theta_R \rightarrow$ \texttt{restrict};  
\quad Else $\rightarrow$ \texttt{approve}.  
If partner $S_j$ involved, enforce cross-domain rules; violations $\rightarrow$ \texttt{deny}.

\State \textbf{Anchoring/Logging:} Record $\{S_i,S_j,R_i,T_i,\text{decision}\}$ in \texttt{GovDecisions} and append to \emph{SORA-Chain} (Table~\ref{tab:sora-policy-matrix}, S2--S3).

\State \textbf{System-level metrics (S4--S6, Def.~7):}  
Compute ecosystem trust and risk. If $\geq 2$ agents exceed high-risk triggers $\rightarrow$ \texttt{joint actuation} (S4). If ecosystem thresholds exceeded $\rightarrow$ \texttt{city-wide escalation} (S5), else apply trust-constrained escalation. Enforce hysteresis $h$ and cooldown $\Delta t$ to maintain stability (S6); anchor outcomes to SORA-Chain.
\end{algorithmic}
\end{algorithm}

\section{Use Case Description: Decentralized, Trusted GRC for Smart-City Disaster Management}
Building on prior work advocating hybrid governance that preserves city-level policy control while decentralizing execution, we implement a smart-city disaster management use case to balance agility and accountability. Agents operate near their data, applying defense-in-depth and providing operator-legible explanations, while a lightweight governance layer enforces policy, ensures provenance, and maintains auditability without central bottlenecks \cite{SIDDIQUI2024, Karim2025}. Three domain agents: Weather, Traffic, and Safety, operate under dual-chain governance: outputs are signed on the \emph{Agentic Blockchain}, and SORA’s decisions are anchored on the \emph{SORA Blockchain}. Each agent follows the unified orchestration and escalation flow in Algorithm~\ref{alg:unified_enforcement} and Table~\ref{tab:sora-policy-matrix}, as detailed in Section~\ref{sec:workflow}.

\subsection{Weather Agent}
\begin{figure}[h]
\centering
\includegraphics[width=\linewidth]{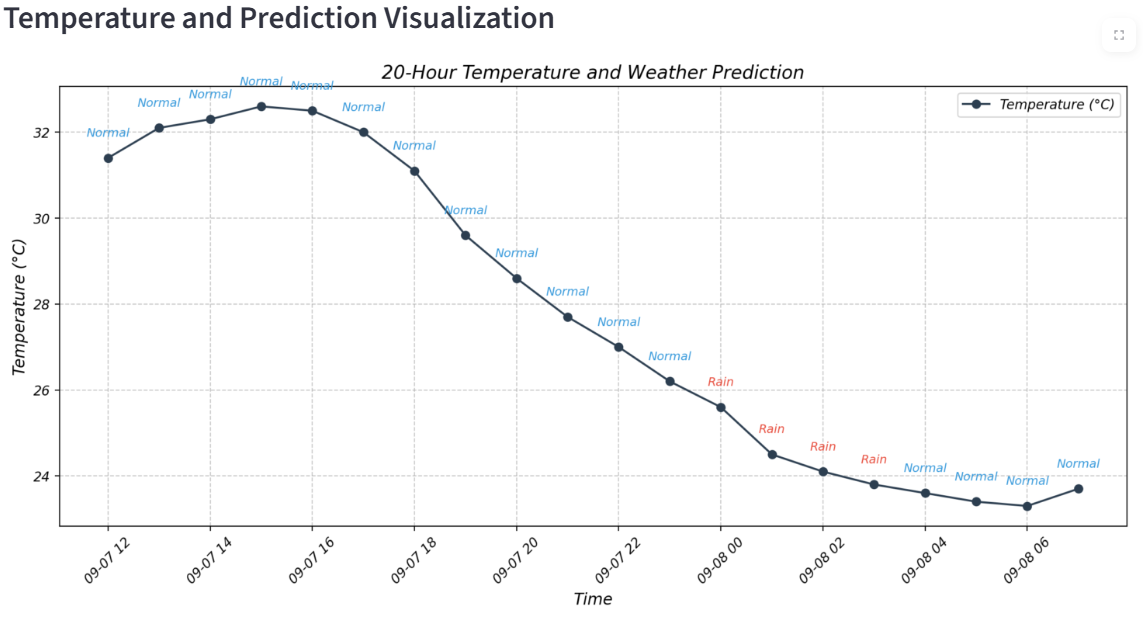}
\caption{20-hour temperature and weather regime forecast (Islamabad example). The line graph shows predicted temperature trends with overlaid regime labels (\emph{Normal}/\emph{Rain}), which are used by the Weather Agent’s trust and risk computations..}
\label{fig:weather-pred}
\end{figure}

The Weather Agent ingests daily and real-time meteorological data for Karachi, Islamabad, and Lahore (2014--2025) from the Open-Meteo API, including temperature, precipitation, humidity, wind speed, cloud cover, UV index, and soil temperature. Following domain literature, we define operational regimes as: \emph{Heavy Rain} ($\geq 40$\,mm/day), \emph{Rain} ($5{-}20$\,mm/day), and \emph{Heatwave} (temperature $\geq 40^\circ$C or anomaly $+5^\circ$C with UV~$\geq8$)~\cite{haseeb2025comprehensive,amjad2022analysis}; otherwise, conditions are labeled \emph{Normal}. These thresholds align with established climate risk criteria for urban flooding, extreme heat exposure, and resilience assessments in South Asian megacities. A supervised XGBoost model trained on the labeled dataset classifies regimes and produces 20-hour forecasts, with performance metrics summarized in Table~\ref{tab:agent-summary}. Forecast outputs are logged in structured CSV files and consumed by three LLMs (ChatGPT, Grok, DeepSeek), which generate risk-trust narratives and signed outputs on the \emph{Agentic Blockchain}. These regime forecasts (Fig.~\ref{fig:weather-pred}) directly inform contextual trust updates and guide SORA’s policy enforcement.  

\subsection{Traffic Agent}
\begin{figure}[h]
\centering
\includegraphics[width=0.74\linewidth]{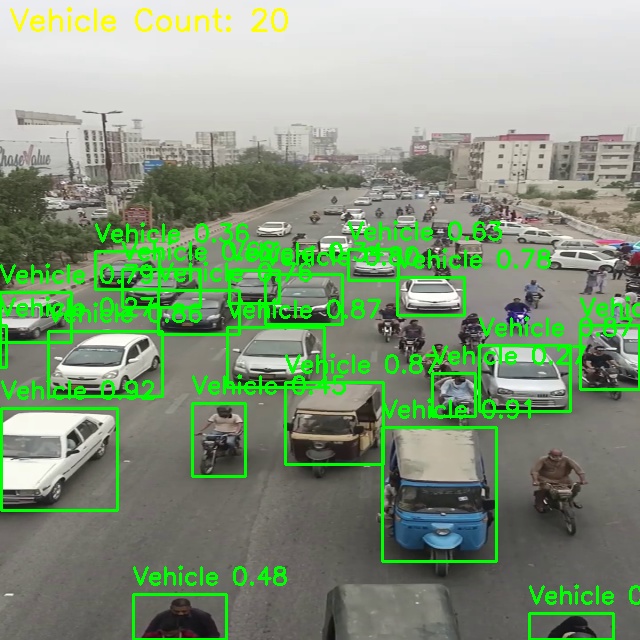}
\caption{YOLOv8 traffic-density output with bounding boxes and confidence scores. Aggregated counts feed directly into the Traffic Agent’s risk-trust pipeline.}
\label{fig:traffic-yolo}
\end{figure}

The Traffic Agent applies a pre-trained YOLOv8 model \cite{YOLOv8_Traffic}, which has been fine-tuned on a custom vehicle dataset, to monitor vehicular density, estimating counts per 100\,m road segment. In line with urban mobility studies, a threshold of \textit{15 vehicles per 100\,m} is adopted as an indicator of congestion \cite{kausar2024evaluating}. The vehicle count is determined by the number of bounding boxes generated by the model within each 100-meter segment, with each bounding box corresponding to a detected vehicle. Model performance is summarized in Table~\ref{tab:agent-summary}. As shown in Fig.~\ref{fig:traffic-yolo}, the model produces per-vehicle bounding boxes, confidence scores, and aggregate counts ("Vehicle Count: 20"), which are logged in structured CSV files. These logs are consumed by the agent’s three LLMs and by SORA to assess congestion risk and reporting trust.

Ground-truth labeling for the dataset was performed through manual annotation of vehicle positions in traffic video footage, ensuring accurate training and evaluation of the model. Recommendations such as diversions or adaptive signaling are signed and anchored on the Agentic Blockchain, ensuring traceable coordination without exposing raw imagery or personally identifiable information.

\subsection{Safety (Fire/Smoke) Agent}
\begin{figure}[h]
\centering
\includegraphics[width=0.82\linewidth]{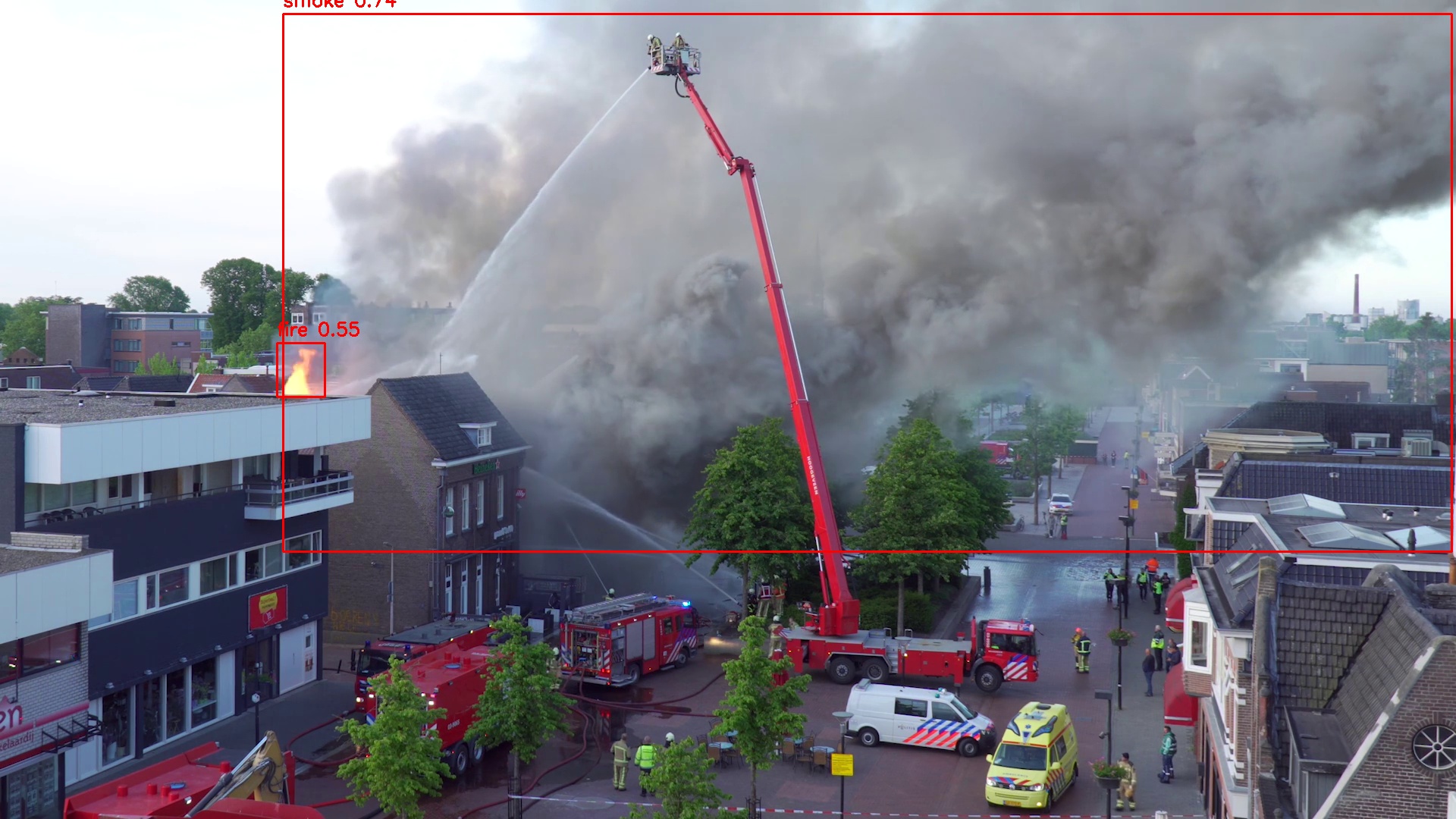}
\caption{YOLO11 "Flare Guard" output on a hazard scene: smoke (0.74) and fire (0.55) regions detected.}
\label{fig:fire-yolo}
\end{figure}

The Safety Agent uses a pre-trained YOLO11 "Flare Guard" model \cite{YOLO11_FireSmoke} to detect smoke and fire events in CCTV streams (performance in Table~\ref{tab:agent-summary}). Each detection, logged with timestamp, type, confidence, and location, is processed by three LLMs for enriched annotations. Signed outputs are anchored on the \emph{Agentic Blockchain}. As shown in Fig.~\ref{fig:fire-yolo}, confidence scores (e.g., smoke 0.74, fire 0.55) inform Environmental Risk and guide SORA’s escalation policies.

\subsection{Governance Authority}
\textbf{SORA} (Security \& Operational Response Agent) acts as the supervisory governance node, maintaining its own trust–risk reference to evaluate each agent’s LLM outputs. The model with the lowest mean absolute error (MAE) per interval is selected as authoritative, while others receive signed, policy-based feedback for convergence. All governance artifacts validations, selections, feedback, policy versions, and directives are immutably recorded on the \emph{SORA Blockchain}, with the \emph{Agentic Blockchain} ensuring edge provenance. This dual-ledger setup supports embedded compliance, human-on-the-loop oversight, and tamper-evident auditability in multi-agent smart-city systems \cite{SIDDIQUI2024, nist2023airmf, smuha2025regulation}.

\begin{table*}[t]
\caption{Overview of domain agents and governance authority with data scope, outputs, performance, and ledger anchoring}
\label{tab:agent-summary}
\small
\centering
    \begin{tabular}{p{2.8cm}p{3.8cm}p{4.5cm}p{4.5cm}}
    \toprule
    \textbf{Agent/Authority} & \textbf{Data Scope} & \textbf{Primary Outputs} & \textbf{Performance \& Ledger} \\
    \midrule
    \textbf{Weather Agent} & 
    Open-Meteo (2014-2025): temp, precip, RH, wind, cloud, UV & 
    Regime classification (Normal, Rain, Heavy Rain, Heatwave); risk/trust rationale & 
    XGBoost (Acc.: train=0.998, test=0.978); \emph{Agentic Chain} \\
    \addlinespace
    
    \textbf{Traffic Agent} & 
    Vehicle counts (YOLOv8): 15 vehicles/100m threshold & 
    Congestion risk/trust; signaling \& diversion recommendations & 
    YOLOv8 (mAP@50=0.975, mAP@50-95=0.742); \emph{Agentic Chain} \\
    \addlinespace
    
    \textbf{Safety Agent} & 
    Smoke/fire detections (YOLO11 "Flare Guard") & 
    Hazard extent; confidence-based annotations & 
    YOLO11 (mAP@50=0.770, mAP@50-95=0.492); \emph{Agentic Chain} \\
    \addlinespace
    
    \textbf{SORA Governance} & 
    Cross-agent oversight \& policy enforcement & 
    MAE-based LLM selection; feedback; global directives & 
    Governance records; \emph{SORA Chain} (dual-ledger) \\
    \bottomrule
    \end{tabular}
\end{table*}

\subsection{Operational Policy Matrix}
Governance enforcement in SORA builds on the trust and risk constructs defined in Section~\ref{metrics}, operating on Overall Agent Risk $R^{i}(t)$ (Def.~4), Overall Trust $T^{i}_{\text{Overall}}(t)$ (Def.~6), and aggregated ecosystem metrics (Def.~7). The unified policy matrix (Table~\ref{tab:sora-policy-matrix}) integrates agent triggers, governance filters, model selection, feedback, and cross-agent escalation, ensuring that advisories, rerouting, or hazard protocols activate only when thresholds are crossed, with stability maintained via hysteresis and cooldown controls.

\begin{table*}[h]
\centering
\small
\renewcommand{\arraystretch}{1.15}
\caption{SORA unified policy matrix: agent triggers, governance filters, and escalation (using $R$, $T$, and ecosystem metrics from Defs.~4--7).}
\label{tab:sora-policy-matrix}
\begin{tabular}{|p{1.6cm}|p{0.9cm}|p{5.9cm}|p{3.0cm}|p{4.0cm}|}
\hline
\textbf{Scope} & \textbf{ID} & \textbf{Trigger / Rule} & \textbf{Primary Action} & \textbf{Notes} \\
\hline
\multicolumn{5}{|c|}{\textbf{Agent-Level Policies}} \\
\hline
Weather & W1 & $R^{\text{Wea}}(t)>0.60 \wedge T^{\text{Wea}}_{\text{Overall}}(t)<0.65$ & Flood/heatwave advisories & Select LLM closest to SORA reference (post~S1) \\
\hline
Traffic & T1 & $R^{\text{Tra}}(t)\ge0.95 \wedge T^{\text{Tra}}_{\text{Overall}}(t)<0.65$ & Rerouting, signal tuning & Select closest-to-SORA LLM (post~S1) \\
\hline
Fire & F1 & $R^{\text{Fire}}(t)\ge0.95 \wedge T^{\text{Fire}}_{\text{Overall}}(t)>0.65$ & Dispatch, evacuation & Enforce most consistent LLM (post~S1) \\
\hline
\multicolumn{5}{|c|}{\textbf{Governance-Level Policies}} \\
\hline
SORA & S1 & Risk–Trust Gate: admit if $\lvert\Delta R\rvert\le0.07$, $\lvert\Delta T\rvert\le0.05$, and $T^{\text{Overall},j}_{i}\ge\tau_T(i)$ (Wea:0.60, Tra:0.55, Fire:0.65) & Admit candidates & Code vars: \texttt{risk\_thresh}=0.07, \texttt{trust\_thresh}=0.05 \\
\hline
SORA & S2 & Selection \& Tie-break: choose $\max T$; if tie, pick $\min\lvert\Delta R\rvert$ (ties within 0.01 equal). Fallback (Fire only): nearest $T\ge\tau_T$, else nearest $T$ & Select best model & No fallback for Weather/Traffic \\
\hline
SORA & S3 & Error-Directed Feedback: $\Delta R=R_{\text{SORA}}-R_j$, $\Delta T=T_{\text{SORA}}-T_j$ & Feedback to non-selected & Apply 50\% adjustment; clip $[0,1]$ \\
\hline
\multicolumn{5}{|c|}{\textbf{Cross-Agent / Ecosystem Policies}} \\
\hline
Cross-Agent & S4 & $\ge2$ agents with $R^{i}(t)>0.80$ & Joint actuation (reroutes, co-alerts) & $T^{i}_{\text{Overall}}$ modulates confidence \\
\hline
Ecosystem & S5 & $R_{\text{Ecosystem}}(t)>0.70 \wedge T_{\text{Ecosystem}}(t)\ge0.60$ & City-wide escalation & Human confirmation if $T_{\text{Ecosystem}}<0.60$ \\
\hline
Safety & S6 & Hysteresis $h=0.05$, cooldown $\Delta t_{\min}=15$\,min & Stability / anti-chatter & Prevent oscillations and alert spam \\
\hline
\end{tabular}

\vspace{4pt}
\footnotesize
\emph{Abbrev.:} Wea=Weather, Tra=Traffic, Fire=Fire(Safety). $R$=Overall Risk (Def.~4), $T$=Overall Trust (Def.~6).
\end{table*}

\begin{figure*}[h]
\centering
\includegraphics[scale=0.32]{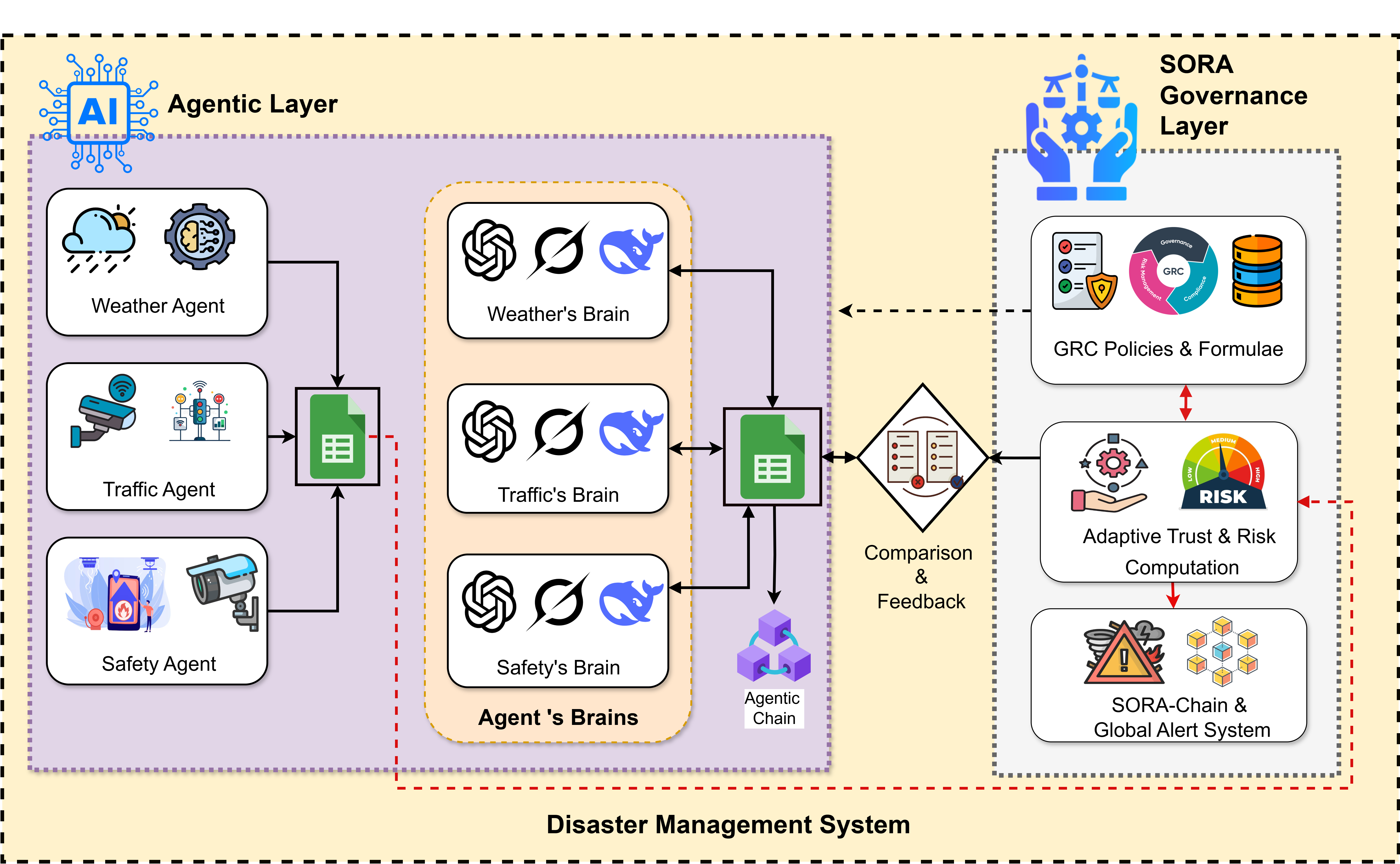}
\caption{End-to-end workflow of the SORA-ATMAS architecture showing API-based orchestration of LLMs, local agentic validation, and global SORA enforcement with dual blockchain anchoring.}
\label{fig:workflow_execution}
\end{figure*}

\section{Workflow Explanation and Execution} \label{sec:workflow}
The workflow integrates decentralized AI agents, LLM-based reasoning, and centralized governance to operationalize adaptive trust management in disaster response. As shown in Fig.~\ref{fig:workflow_execution}, local agents process sensor data, invoke multiple LLMs for semantic reasoning, and generate structured risk–trust outputs that are relayed to the SORA Governance Layer for validation, compliance enforcement, and escalation. Unlike static pipelines, this workflow supports context-aware, risk-sensitive decisions across both stable and high-risk scenarios. The closed feedback loop $(\Delta R,\Delta T)$ iteratively returned to agent nodes drives convergence and MAE reduction (Section~\ref{sec:results}). The setup was implemented on a distributed testbed (as shown in Table~\ref{tab:hardware_config}).

\paragraph{\textbf{Agent-Level Workflow (Local Processing on PC-A)}} 
\begin{enumerate}
    \item Each domain-specific agent (Weather, Traffic, Safety) collects contextual sensor data (e.g., temperature, congestion levels, fire/smoke frames) and records it into structured logs on Google Sheets. A sample message structure is defined as:
    \begin{equation}
    M = \{ \text{AID}, D, t, \lambda \},
    \label{eq:agent_message}
    \end{equation}
    where $\text{AID}$ denotes the agent identifier, $D$ represents the sensor data payload, $t$ is the timestamp, and $\lambda$ encodes local compliance metadata. 

    \item The logged entries are transmitted via API calls to multiple LLMs (ChatGPT, Grok, DeepSeek), which act as the semantic reasoning “brains.” Each model computes domain-specific trust and risk components:
    \begin{equation}
    \{R^i_{\text{Env}},\ R^i_{\text{Service}},\ T^i_{\text{Ctx}},\ T^i_{\text{HRT}}\},
    \end{equation}
    which are then aggregated into overall trust–risk scores $R^i(t)$ and $T^i(t)$ according to the definitions in Section~\ref{metrics}. 

   \item The agent signs and forwards the per-LLM outputs $\{R^i(t),T^i(t)\}$ to SORA. The signed packet is anchored on the Agentic Blockchain for provenance.
  
\end{enumerate}

\paragraph{SORA Governance Workflow (on PC-B)} 
\begin{enumerate}
    \item At the governance node, SORA first ingests raw sensor values transmitted by the agents. These inputs are checked using the policies given by the Security Policy Engine for authentication, authorization, and access control and by the Cross-Domain Operational Policy Engine for domain-specific compliance. Only packets that pass both checks are admitted, and from these, SORA computes the governance-level reference risk $R^{\text{ref}}$ and trust $T^{\text{ref}}$ values according to the definitions in Section~\ref{metrics}, which form the baseline for comparison.  
    \item SORA then receives the signed per-LLM outputs $\{R^i(t),T^i(t)\}$ from each agent. These packets are again verified for provenance, authorization, and compliance. Invalid or non-compliant packets are discarded before further processing.  

    \item For each valid packet, SORA calculates the Mean Absolute Error (MAE) between an LLM’s reported outputs 
    and its governance reference values:
   \begin{equation}
        \text{MAE}_j = \frac{1}{2}\Big(|R_j - R^{\text{ref}}| + |T_j - T^{\text{ref}}|\Big),
    \label{eq:mae}
    \end{equation}

    where $j$ indexes the LLMs. These MAE scores are used in policy step S2 (model selection) and step S3 (error-directed feedback).  

    \item The Adaptive Trust and Risk Enforcement Engine applies global thresholds as specified in the policy matrix. Formally:
     \begin{equation}
    \texttt{Decision} =
    \begin{cases}
    \texttt{deny}, & T^i < \theta_T, \\
    \texttt{restrict}, & R^i > \theta_R, \\
    \texttt{approve}, & \text{otherwise}.
    \end{cases}
    \label{eq:decision}
    \end{equation}

    This ensures that high-risk or low-trust outputs cannot bypass systemic safeguards.  

    \item Cross-agent and ecosystem-level policies are then enforced. If at least two agents report high risk ($R^i>0.80$), 
    joint actuation is triggered; if the ecosystem risk exceeds $0.70$ with ecosystem trust $\ge 0.60$, 
    City-wide escalation follows (with human confirmation required if trust is lower).  

    \item All governance decisions, including model selection, feedback, joint actions, and escalation outcomes, are immutably anchored on the SORA Blockchain. In addition, all dispatched alerts (e.g., flood advisories, traffic reroutes, fire evacuations) are also logged on-chain for provenance and post-hoc audit, while being simultaneously delivered to stakeholders via API channels.
  
\end{enumerate}

This layered synchronization between PC-A and PC-B is maintained through structured data exchange (Google Sheets API), 
tamper-resistant anchoring on the Agentic and SORA Blockchains, and lightweight messaging for timely dissemination of alerts and directives.

\section{Results and Discussion} 
\label{sec:results}
The evaluation of the SORA-ATMAS framework is conducted along two complementary dimensions: semantic alignment of local LLMs with governance-enforced reference outputs, and operational performance of SORA as a centralized authority within a distributed multi-agent setting including cross-domain interoperability for coordinated escalation. On the semantic side, the focus lies on how candidate models progressively converge toward policy-aligned outputs under governance supervision. On the operational side, we analyze throughput, execution time, and governance-induced delay across varying workloads to capture the trade-offs between oversight, responsiveness, and blockchain anchoring. Together, these results demonstrate SORA’s ability to balance correctness, safety, and scalability in smart-city disaster governance.

\subsection{Agent-Level Evaluation of LLMs}
Three domain agents (Weather, Traffic, and Fire/Smoke) were evaluated across three iterations ($\text{Iter.}~0$–$2$). Each agent processed 20 structured requests per iteration, yielding 60 total observations per model per iteration. This smaller, controlled batch size was used for semantic convergence analysis rather than large-scale throughput testing. The selection of candidate LLMs (GPT, Grok, DeepSeek) was governed by two joint criteria: 

\begin{enumerate}
    \item Minimizing the Mean Absolute Error (MAE) with respect to SORA’s reference values, and 
    \item Satisfying governance policies (S1-S3) as given in Table ~\ref{tab:sora-policy-matrix}. 
\end{enumerate}

\begin{figure*}[t]
\centering
\includegraphics[scale=0.24]{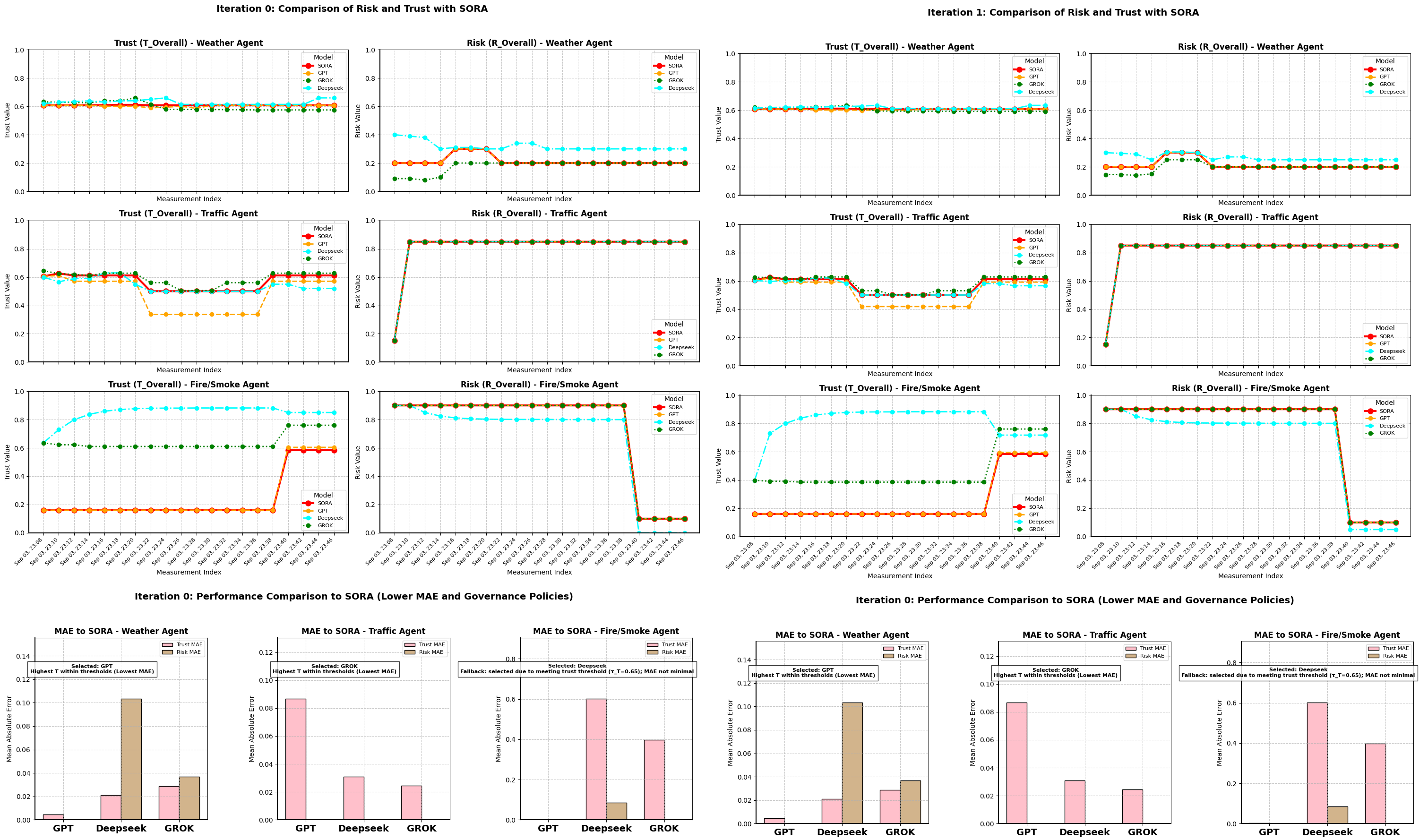}
\caption{Iterations~0 and~1: Comparison of candidate LLMs with SORA across Weather, Traffic, and Fire/Smoke agents. 
Top: Trust and Risk trajectories. Bottom: MAE-to-SORA distributions, showing initial selection and the first reduction in MAE after feedback.}
\label{fig:iter0}
\end{figure*}

\begin{figure*}[h]
\centering
\includegraphics[scale=0.3]{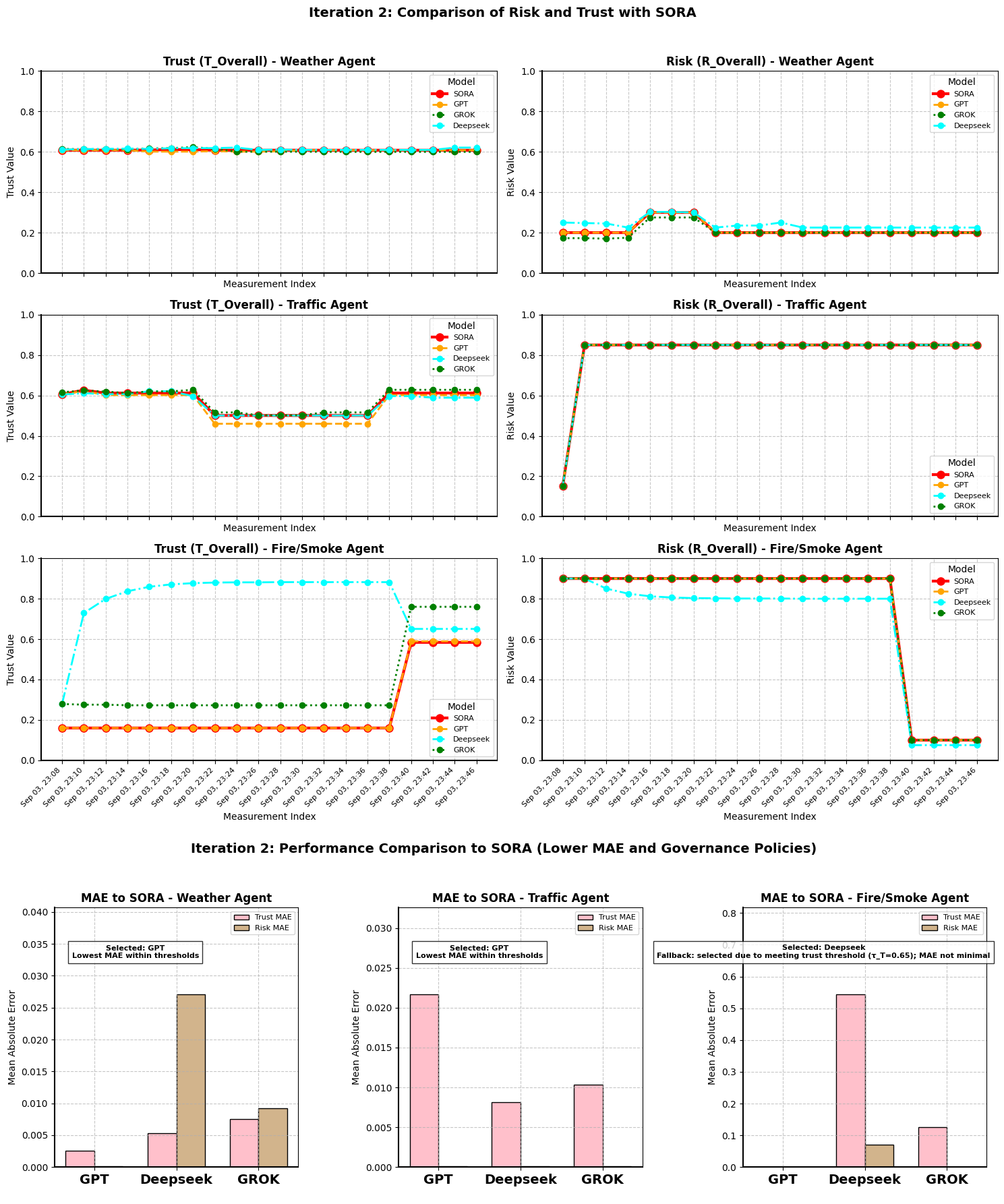}
\caption{Iteration~2: Convergence of candidate LLMs toward SORA with stable risk alignment and further reduction in trust MAE.}
\label{fig:iter2}
\end{figure*}

Specifically, the policies enforced a $\texttt{risk\_threshold}=0.07$, a $\texttt{trust\_threshold}=0.05$, and domain-specific trust baselines $\tau_t=\{0.60,0.55,0.65\}$ for Weather, Traffic, and Fire/Smoke, respectively.  When no candidate satisfied S1, the fallback mechanism was activated, allowing a model to be selected solely for meeting the trust baseline $\tau_t$, even if its MAE was not minimal. Error-directed feedback $(\Delta R,\Delta T)$ was iteratively applied with an adjustment factor of $0.5$ (clipped to $[0,1]$), 
driving non-selected models toward convergence. Figures~\ref{fig:iter0}-\ref{fig:iter2} show the trajectory of risk-trust values and the evolution of MAE distributions.

\paragraph{Weather Agent:} The Weather agent maintained low $R_{\text{Overall}}$ values across iterations. At $\text{Iter.}~0$, GPT achieved the lowest MAE (risk and trust). By $\text{Iter.}~1$, error feedback allowed DeepSeek to significantly reduce its trust MAE, and by $\text{Iter.}~2$, DeepSeek surpassed GPT with the lowest trust MAE while maintaining stable risk alignment. Risk MAE remained negligible, confirming stable conditions.

\paragraph{Traffic Agent:} Traffic data showed a high-risk plateau ($R\approx0.85$). Grok consistently achieved the lowest MAE across iterations, dominating in $\text{Iter.}~0$ and $\text{Iter.}~1$. GPT and DeepSeek reduced their trust MAE over iterations, as seen in Figures~\ref{fig:iter0}–\ref{fig:iter2}. By $\text{Iter.}~2$, Grok retained dominance, illustrating its stability under congestion. Risk MAE remained near zero, making trust MAE the decisive factor.

\paragraph{Safety (Fire/Smoke) Agent:} The Fire/Smoke agent transitioned from high-risk/low-trust ($R>0.9, T<0.2$) to low-risk/high-trust ($R\approx0.1, T>0.6$). At $\text{Iter.}~0$, most models failed the S1 trust gate ($\tau_T=0.65$), activating the fallback policy, which selected DeepSeek despite its higher MAE. By $\text{Iter.}~1$, DeepSeek maintained dominance while lowering its trust MAE. By $\text{Iter.}~2$, more models satisfied S1, but DeepSeek still outperformed with the lowest trust MAE.

\paragraph{Key Trends:} Across all agents, SORA’s governance (S1 gating, S2 tie-breaking, S3 feedback, fallback) consistently reduced trust MAE while keeping risk MAE stable. These results demonstrate that governance not only ensured threshold compliance but also minimized MAE, guiding models toward SORA’s reference behavior while maintaining safety in dynamic conditions. These semantic alignments at the agent level ultimately enable coherent cross-domain reasoning, allowing SORA to merge independent risk–trust assessments into synchronized, policy-aligned escalation actions.

\subsection*{MAE Reductions and Statistical Analysis}
\begin{figure*}[h]
\centering
\includegraphics[width=0.85\linewidth]{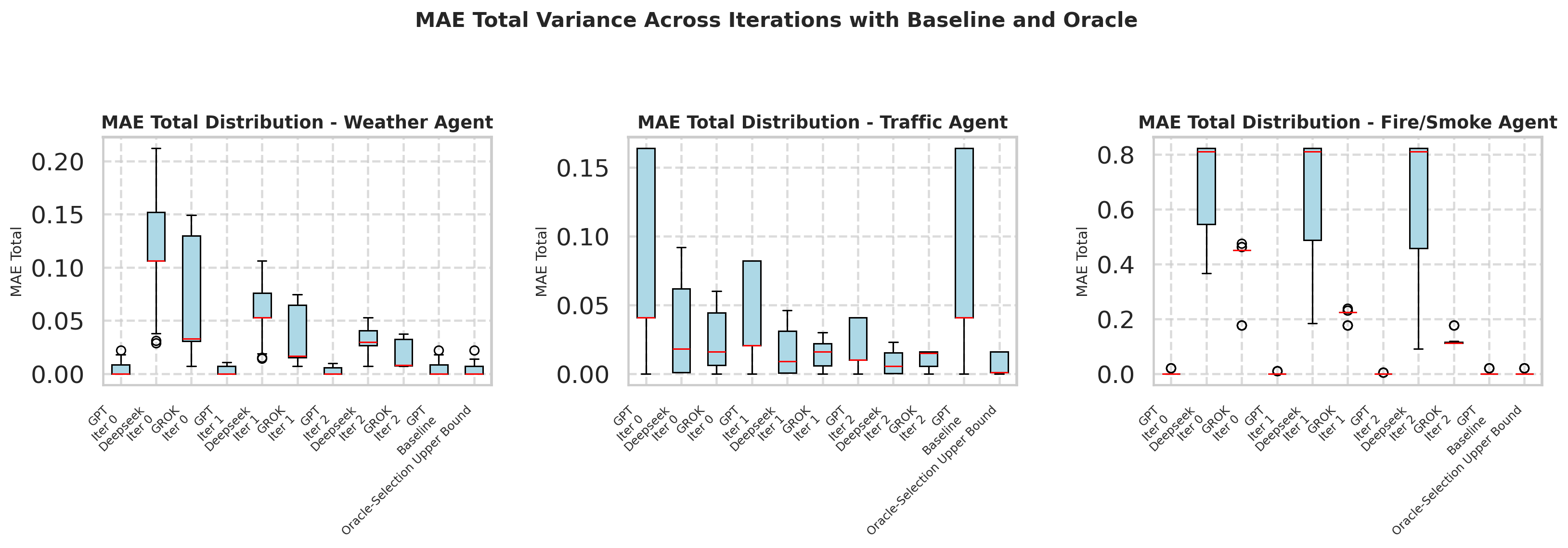}
\caption{MAE distribution across iterations for Weather, Traffic, and Fire/Smoke agents, showing reductions per model with baseline (GPT) and oracle upper bound for comparison.}
\label{fig:boxplot-iter0}
\end{figure*}

The MAE reductions across iterations as shown in Figure ~\ref{fig:iter0}, and Figure ~\ref{fig:iter2} were analyzed using boxplots and Wilcoxon Signed-Rank tests. The boxplots figure ~\ref{fig:boxplot-iter0} show the variance in MAE Total reductions across iterations for the Weather, Traffic, and Fire/Smoke agents. These reductions are compared against the \emph{single-LLM baseline (GPT)} and the \emph{oracle-selection upper bound} to assess model performance.

The Wilcoxon Signed-Rank test results for MAE reductions across iterations are presented in Table~\ref{tab:wilcoxon-mae}. P-values below 0.05 indicate statistically significant reductions in MAE. Except for GPT in certain agent iterations (e.g., Weather Agent), all models show significant reductions.

\begin{table}[h]
\centering
\caption{Wilcoxon Signed-Rank Test Results for MAE Reductions (p-values for Iter 0$\to$1 / Iter 1$\to$2)}
\begin{tabular}{|c|c|c|}
\hline
\textbf{Agent} & \textbf{Model} & \textbf{P-Values} \\
\hline
Weather & GPT       & 0.1088 / 0.1797 \\
Weather & DeepSeek  & 0.000078 / 0.000120 \\
Weather & GROK      & 0.000124 / 0.000124 \\
Traffic & GPT       & 0.000087 / 0.000087 \\
Traffic & DeepSeek  & 0.000077 / 0.000166 \\
Traffic & GROK      & 0.006459 / 0.002968 \\
Fire/Smoke & GPT    & 0.0455 / 0.0455 \\
Fire/Smoke & DeepSeek & 0.0339 / 0.0339 \\
Fire/Smoke & GROK   & 0.000175 / 0.000175 \\
\hline
\end{tabular}
\label{tab:wilcoxon-mae}
\end{table}

Additionally, the Mean MAE for the single-LLM baseline (GPT) and the oracle-selection upper bound are provided in Table~\ref{tab:mean-mae-baseline}. The single-LLM baseline (GPT) is used as a comparison to assess model performance, and the oracle-selection upper bound represents the best possible model selection.

\begin{table}[h]
\centering
\caption{Mean MAE for Single-LLM Baseline and Oracle-Selection Upper Bound}
\begin{tabular}{|c|c|}
\hline
\textbf{Agent} & \textbf{Mean MAE} \\
\hline
Weather & 0.00445 (GPT), 0.0039 (Oracle-Selection) \\
Traffic & 0.08685 (GPT), 0.0072 (Oracle-Selection) \\
Fire/Smoke & 0.00400 (GPT), 0.0040 (Oracle-Selection) \\
\hline
\end{tabular}\label{tab:mean-mae-baseline}
\end{table}

\subsection{Performance Evaluation of SORA}
\label{sec:sora-perf}
After establishing agent-level semantic alignment, we evaluate SORA’s runtime performance as a governance authority. Beyond local processing, SORA validates reports, enforces policies, and anchors decisions on the blockchain, introducing governance overhead measured by three key metrics:

\begin{itemize}
    \item \textbf{Throughput ($T$):} The effective rate of governance-approved requests per second:
    
    \begin{equation}
        T = \frac{N}{t^{\text{last decision}} - t^{\text{first request}}},
    \end{equation}
    where $N$ is the number of processed requests.
    
    \item \textbf{Execution Time (ET):} The average per-request latency at the agent level, comprising policy fetching, local risk–trust computation, and secure logging to the Agent-Chain:
    \begin{equation}
        \text{ET} = t^{\text{fetch}} + t^{\text{compute}} + t^{\text{Agentic-Chain log}} .
    \end{equation}
    
    \item \textbf{Operational Delay ($D$):} The additional per-request overhead introduced by SORA governance, consisting of global validation, MAE-based LLM selection, corrective feedback dissemination, final decision synthesis, and SORA-Chain anchoring:
    \begin{equation}
        D = t^{\text{validate}} + t^{\text{MAE-select}} + t^{\text{feedback}} + t^{\text{final-decision}} + t^{\text{SORA-Chain log}} .
    \end{equation}
\end{itemize}

\paragraph{Experimental Testbed Configuration:} The performance evaluation was conducted on a distributed two-layer testbed consisting of multiple agent nodes and a single governance node. The hardware, network, and blockchain configurations used for experiments are summarized in Table~\ref{tab:hardware_config}.

\begin{table}[h]
\centering
\caption{Hardware configuration for SORA evaluation.}
\label{tab:hardware_config}
\small
\setlength{\tabcolsep}{2pt}
\renewcommand{\arraystretch}{1.05}
\begin{tabular}{|p{2.5cm}|p{5.5cm}|}
\hline
\textbf{Component} & \textbf{Specifications} \\
\hline
Agent Nodes (3× PC-A) & Intel\textsuperscript{\textregistered} i7-12700K (3.8\,GHz, 32 threads); 32\,GB RAM; 1\,TB SSD; Ubuntu 22.04. \\
\hline
Governance Node (PC-B) & AMD Ryzen 9 5900X (64 threads); 64\,GB RAM; 2\,TB SSD; Ubuntu 22.04. \\
\hline
Network & Gigabit LAN, $<$2\,ms latency. \\
\hline
Blockchain & Multichain: synchronous (SORA-Chain), asynchronous/batched (Agentic-Chain). \\
\hline
\end{tabular}
\end{table}

\paragraph{Experimental Setup:}
The performance evaluation used synthetic workloads simulating structured observations from domain agents such as Weather, Traffic, and Safety, with request sizes of 100, 500, 1000, and 2000. To assess scalability, a fixed workload of 100 requests was used to measure throughput (T), execution time (ET), and governance delay (D) across different agent configurations. Larger configurations with six and nine agents were projected using an analytical M/M/c queueing model~\cite{chaudhary2025ai} calibrated to the measured three-agent results, as hardware constraints limited physical scalability beyond three nodes. Timers at both the agent and governance layers recorded ET and D, while T was derived from the total wall-clock duration.

\paragraph{Runtime Performance and Scalability of SORA:}
The runtime performance of SORA, including blockchain anchoring and decision synthesis, was measured using synthetic workloads of varying sizes. For the 3-agent configuration, results for throughput, execution time (ET), and governance delay (D) are summarized in Table~\ref{tab:sora-perf}.

\begin{table}[h]
\centering
\small
\caption{Runtime performance of SORA including blockchain anchoring and decision synthesis.}
\label{tab:sora-perf}
\setlength{\tabcolsep}{5pt}
\renewcommand{\arraystretch}{1.15}
\begin{tabular}{rcccc}
\toprule
\textbf{Requests} & \textbf{$T$ (req/s)} & \textbf{ET / req (ms)} & \textbf{$D$ / req (ms)} \\
\midrule
100  & 17.2 & 58 & 21 \\
500  & 16.3 & 61 & 32 \\
1000 & 15.2 & 65 & 52 \\
2000 & 13.8 & 72 & 92 \\
\bottomrule
\end{tabular}

\vspace{1mm}
\footnotesize
\raggedright
\textbf{Notes:} $T$ = Throughput, ET = Execution Time per request, and $D$ = Operational/Governance Delay per request.
\end{table}

\begin{figure}[h]
\centering
\includegraphics[width=0.45\textwidth]{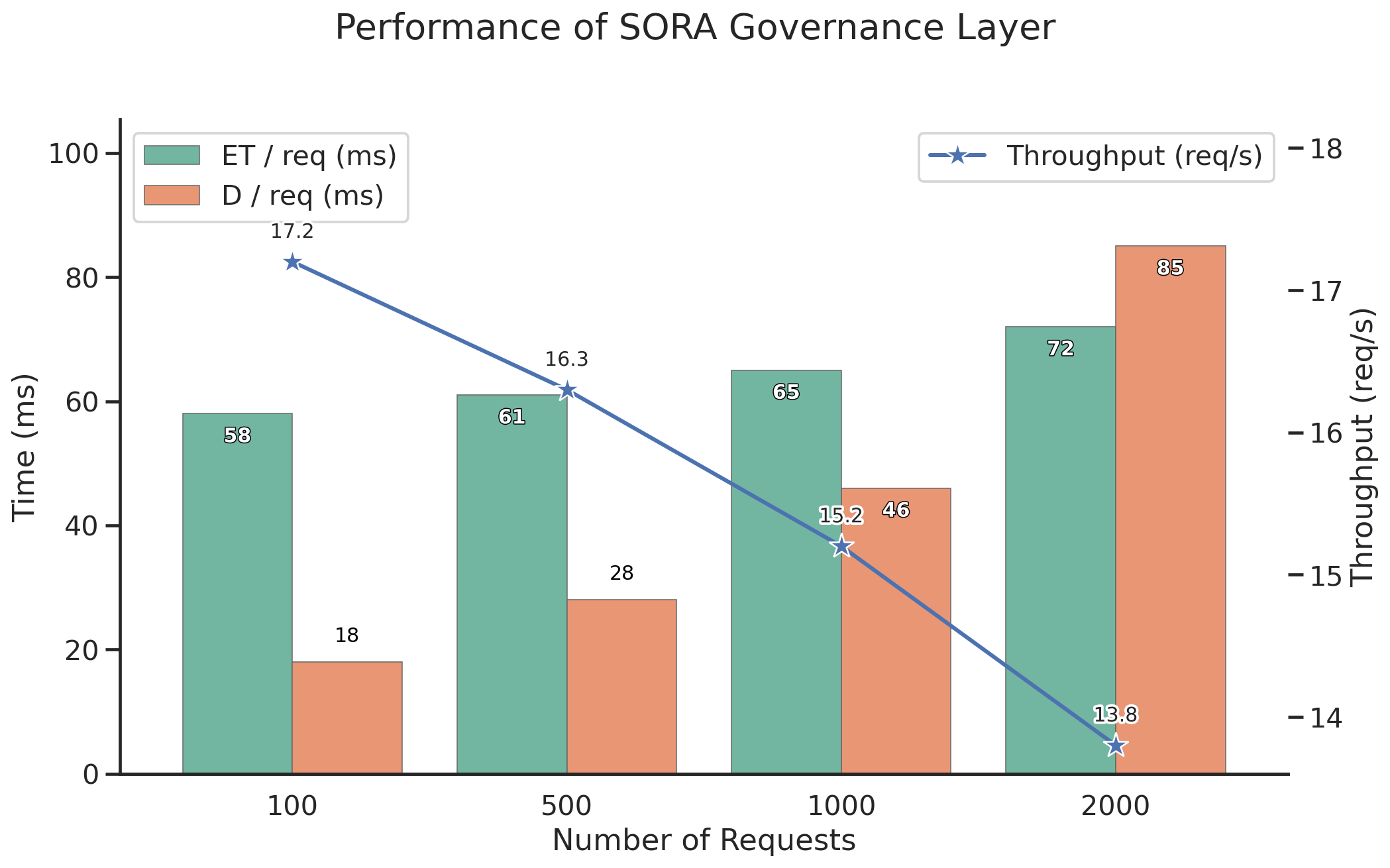}
\caption{Performance of SORA governance layer for the 3-agent deployment (ET and governance delay shown as bars; throughput plotted as line).}
\label{fig:sora-runtime}
\end{figure}

The scalability of SORA was further analyzed across different agent configurations (3, 6, and 9 agents). For the 3-agent configuration, we measured throughput, execution time, and governance delay. For the 6- and 9-agent configurations, the results were projected using an analytical M/M/c queueing model, as shown in Table~\ref{tab:sora-perf-scalability}.

\begin{table}[h]
\centering
\caption{Scalability analysis across different agent configurations. Projections are simulation-based on fixed 100 requests.}
\label{tab:sora-perf-scalability}
\small
\setlength{\tabcolsep}{4pt}
\renewcommand{\arraystretch}{1.1}
\begin{tabular}{lccc}
\toprule
\textbf{Metric} & \textbf{\shortstack{3 Agents\\(Measured)}} & \textbf{\shortstack{6 Agents\\(Projected)}} & \textbf{\shortstack{9 Agents\\(Projected)}} \\
\midrule
\textbf{$T$ (req/s)} & 17.2 & 15.8 & 14.1 \\
\textbf{ET / req (ms)} & 58 & 64 & 71 \\
\textbf{$D$ / req (ms)} & 21 & 38 & 62 \\
\bottomrule
\end{tabular}

\vspace{1mm}
\footnotesize
\raggedright
\textbf{Notes:} $T$ = Throughput, ET = Execution Time per request, and $D$ = Operational/Governance Delay per request.
\end{table}

\paragraph{Throughput Trends:}
As shown in Table~\ref{tab:sora-perf} and Figure~\ref{fig:sora-runtime}, throughput declines moderately with workload size, from 17.2~req/s at 100 requests to 13.8~req/s at 2000, primarily due to serial governance operations and dual-chain anchoring. Throughput remains above 13~req/s even under the maximum load, confirming practical scalability for near-real-time applications.

\paragraph{Execution Time Analysis:}
Execution time at the agent level increases modestly (58~ms to 72~ms) with higher workloads, indicating that policy fetching, local risk–trust computation, and Agentic-Chain logging scale efficiently on CPU-only agent nodes. This confirms the lightweight nature of local agent pipelines and efficient blockchain integration.

\paragraph{Operational Delay Trends:}
Governance/Operational delay ($D$) rises more sharply (21~ms to 92~ms) as workloads grow, reflecting the additional costs of validation, model selection, and synchronous SORA-Chain anchoring. Despite this growth, delay remains under 100~ms for the tested range, supporting the feasibility of responsive governance at modest agent scales.

\paragraph{Scalability Discussion:}
The analytical projections in Table~\ref{tab:sora-perf-scalability} extend the measured results to larger deployments. Under equivalent per-agent load and network conditions, SORA can sustain throughput above 14~req/s with nine agents. However, governance delay grows substantially with increased coordination and synchronous anchoring overhead. These findings identify the governance layer as the principal scaling constraint, motivating future enhancements such as SORA sharding, partial asynchrony, and multi-chain partitioning to improve scalability in city-scale deployments.

Overall, SORA achieves a practical balance between governance oversight and runtime responsiveness. The distributed testbed (Table~\ref{tab:hardware_config}) demonstrates that CPU-only agent nodes can deliver efficient execution and verifiable blockchain anchoring. The combination of empirical 3-agent measurements and simulation-based projections provides a conservative but realistic view of SORA’s scalability, with clear optimization directions for large-scale smart-city integrations.

\section{Conclusion and Future Directions}
In conclusion, this work introduced \emph{SORA-ATMAS}, a principled governance framework integrating decentralized agentic intelligence with centralized oversight and dual-chain anchoring for resilient smart-city disaster management. By embedding governance and escalation policies (S1–S6), the framework enables heterogeneous agents (Weather, Traffic, and Safety/Fire) to operate autonomously while remaining accountable to city-wide policies. Evaluation showed that multiple LLMs (GPT, Grok, DeepSeek), including SORA’s fallback mechanism for high-risk scenarios, converged toward policy-aligned baselines, with domain-specific optimization (Grok in Traffic, DeepSeek in Weather/Safety) reducing trust deviation (MAE) by $\approx35\%$ while maintaining stable risk alignment. Throughput between 13.8–17.2~req/s, execution times of 58–72~ms, and governance delays of 21–92~ms confirm real-time oversight feasibility under load. Policy-driven coordination further prevented unsafe or conflicting actions, such as traffic rerouting constrained by weather and Fire/Smoke validations. Overall, \emph{SORA-ATMAS} provides a regulation-aligned, verifiable, and context-aware framework for transforming distributed agent outputs into accountable, city-scale decisions.

Future research must now focus on hardening the framework against privacy risks, security threats, and adversarial conditions while further extending its scalability and applicability. Privacy-preserving analytics such as differential privacy, homomorphic encryption, and secure enclaves should be integrated to safeguard inter-domain data exchanges, complemented by selective disclosure mechanisms like zero-knowledge proofs. Strengthened cryptographic assurances through post-quantum primitives, hierarchical key rotation, and formally verified contracts will further reinforce auditability and compliance. Resilience against adversarial scenarios, including data poisoning, sensor spoofing, and LLM prompt injection, will require embedding uncertainty quantification and adversarial detection into both agent-level and supervisory layers. In parallel, static enforcement thresholds can evolve into adaptive, learning-based controllers that preserve safety invariants while adapting to context and operator input. Building on the runtime evaluation and scalability projections presented in Section~\ref{sec:sora-perf}, future work should incorporate large-scale stress tests beyond nine-agent simulations, exploring SORA sharding, partial asynchrony, and multi-chain partitioning to ensure sustained throughput and bounded governance delays in city and multi-city deployments. Scaling SORA-ATMAS with federated policy learning, digital twin–based simulations, and standardized audit trails will provide the operational validation and regulatory grounding necessary for deployment. Advancing along these directions will consolidate SORA-ATMAS as a scalable, privacy-preserving, and resilient foundation for governance in next-generation smart cities.

%\bibliographystyle{elsarticle-harv}
%\bibliographystyle{elsarticle-num} 
%\bibliography{refrences}

\begin{thebibliography}{10}
\expandafter\ifx\csname url\endcsname\relax
  \def\url#1{\texttt{#1}}\fi
\expandafter\ifx\csname urlprefix\endcsname\relax\def\urlprefix{URL }\fi
\expandafter\ifx\csname href\endcsname\relax
  \def\href#1#2{#2} \def\path#1{#1}\fi

\bibitem{rathore2016urban}
M.~M. Rathore, A.~Ahmad, A.~Paul, S.~Rho, Urban planning and building smart cities based on the internet of things using big data analytics, Computer networks 101 (2016) 63--80.

\bibitem{sarker2022smart}
I.~H. Sarker, Smart city data science: Towards data-driven smart cities with open research issues, Internet of Things 19 (2022) 100528.

\bibitem{adeleke2025comprehensive}
O.~J. Adeleke, K.~Jovanovich, S.~Ogunbunmi, O.~Samuel, T.~O. Kehinde, Comprehensive exploration of smart cities: A systematic review of benefits, challenges, and future directions in telecommunications and urban development, IEEE Sensors Reviews (2025).

\bibitem{bittencourt2024survey}
J.~C.~N. Bittencourt, D.~G. Costa, P.~Portugal, F.~Vasques, A survey on adaptive smart urban systems, IEEE Access Volume: 12 (2024).

\bibitem{li2022big}
X.~Li, H.~Liu, W.~Wang, Y.~Zheng, H.~Lv, Z.~Lv, Big data analysis of the internet of things in the digital twins of smart city based on deep learning, Future Generation Computer Systems 128 (2022) 167--177.

\bibitem{al2025smart}
M.~Al-Raeei, The smart future for sustainable development: Artificial intelligence solutions for sustainable urbanization, Sustainable development 33~(1) (2025) 508--517.

\bibitem{tiwari2025agentic}
A.~Tiwari, Conceptualising the emergence of agentic urban ai: from automation to agency, Urban Informatics 4~(13) (2025).
\newblock \href {https://doi.org/10.1007/s44212-025-00079-7} {\path{doi:10.1007/s44212-025-00079-7}}.

\bibitem{murugesan2025rise}
S.~Murugesan, The rise of agentic ai: implications, concerns, and the path forward, IEEE Intelligent Systems 40~(2) (2025) 8--14.

\bibitem{Acharya2025}
D.~B. Acharya, K.~Kuppan, B.~Divya, Agentic ai: Autonomous intelligence for complex goals: A comprehensive survey, IEEE Access 13 (2025) 18912--18936.
\newblock \href {https://doi.org/10.1109/ACCESS.2025.3532853} {\path{doi:10.1109/ACCESS.2025.3532853}}.

\bibitem{HOSSEINI2025}
S.~Hosseini, H.~Seilani, The role of agentic ai in shaping a smart future: A systematic review, Array 26 (2025) 100399.
\newblock \href {https://doi.org/https://doi.org/10.1016/j.array.2025.100399} {\path{doi:https://doi.org/10.1016/j.array.2025.100399}}.

\bibitem{homsy2019multilevel}
G.~C. Homsy, Z.~Liu, M.~E. Warner, Multilevel governance: Framing the integration of top-down and bottom-up policymaking, International Journal of Public Administration 42~(7) (2019) 572--582.

\bibitem{LARTEY2025}
K.~M.~L. Desmond~Lartey, Artificial intelligence adoption in urban planning governance: A systematic review of advancements in decision-making, and policy making, Landscape and Urban Planning 258 (2025) 105337.
\newblock \href {https://doi.org/https://doi.org/10.1016/j.landurbplan.2025.105337} {\path{doi:https://doi.org/10.1016/j.landurbplan.2025.105337}}.

\bibitem{homaei2024}
M.~Homaei, {\'O}.~Mogollón-Guti{\'e}rrez, J.~Sancho, et~al., A review of digital twins and their application in cybersecurity based on artificial intelligence, Artificial Intelligence Review 57~(201) (july 2024).
\newblock \href {https://doi.org/10.1007/s10462-024-10805-3} {\path{doi:10.1007/s10462-024-10805-3}}.

\bibitem{Raj2023}
V.~S. Rajkumar, A.~Ştefanov, A.~Presekal, P.~Palensky, J.~L.~R. Torres, Cyber attacks on power grids: Causes and propagation of cascading failures, IEEE Access 11 (2023) 103154--103176.
\newblock \href {https://doi.org/10.1109/ACCESS.2023.3317695} {\path{doi:10.1109/ACCESS.2023.3317695}}.

\bibitem{Singhal2024}
N.~Singhal, S.~Goyal, T.~Singhal, Emerging Trends in Regulatory and Legal Frameworks for Decentralized Insurance, Springer Nature Singapore, Singapore, 2024, pp. 1--94.
\newblock \href {https://doi.org/10.1007/978-981-97-5894-4_5} {\path{doi:10.1007/978-981-97-5894-4_5}}.

\bibitem{Grace2025}
B.~N. Jørgensen, Z.~G. Ma, Impact of eu laws on the adoption of ai and iot in advanced building energy management systems: A review of regulatory barriers, technological challenges, and economic opportunities, Buildings 15~(13) (2025).
\newblock \href {https://doi.org/10.3390/buildings15132160} {\path{doi:10.3390/buildings15132160}}.

\bibitem{smuha2025regulation}
N.~A. Smuha, Regulation 2024/1689 of the eur. parl. \& council of june 13, 2024 (eu artificial intelligence act), International Legal Materials (2025) 1--148.

\bibitem{sanchez2025ethical}
T.~W. Sanchez, M.~Brenman, X.~Ye, The ethical concerns of artificial intelligence in urban planning, Journal of the American Planning Association 91~(2) (2025) 294--307.

\bibitem{PRAHARAJ2025}
S.~Praharaj, Command and control governance in the 100 smart cities mission in india: Urban innovation or utopias?, Applied Geography 184 (2025) 103766.
\newblock \href {https://doi.org/https://doi.org/10.1016/j.apgeog.2025.103766} {\path{doi:https://doi.org/10.1016/j.apgeog.2025.103766}}.

\bibitem{Karim2025}
M.~M. Karim, D.~H. Van, S.~Khan, Q.~Qu, Y.~Kholodov, Ai agents meet blockchain: A survey on secure and scalable collaboration for multi-agents, Future Internet 17~(2) (2025).
\newblock \href {https://doi.org/10.3390/fi17020057} {\path{doi:10.3390/fi17020057}}.

\bibitem{RIAD2021}
K.~Riad, J.~Cheng, Adaptive xacml access policies for heterogeneous distributed iot environments, Information Sciences 548 (2021) 135--152.
\newblock \href {https://doi.org/https://doi.org/10.1016/j.ins.2020.09.051} {\path{doi:https://doi.org/10.1016/j.ins.2020.09.051}}.

\bibitem{rosmaninho2025}
R.~Rosmaninho, D.~Raposo, P.~Rito, S.~Sargento, Edge-cloud continuum orchestration of critical services: A smart-city approach, IEEE Transactions on Services Computing (2025).
\newblock \href {https://doi.org/10.1109/TSC.2025.3568251} {\path{doi:10.1109/TSC.2025.3568251}}.

\bibitem{musik2025}
S.~Musik, J.~Sasin-Kurowska, M.~Panczyk, Bridging the past and future of clinical data management: The transformative impact of artificial intelligence, Open Access Journal of Clinical Trials 17 (2025) 15--33.

\bibitem{islam2025trust}
R.~Islam, R.~Bose, S.~Roy, et~al., Decentralized trust framework for smart cities: a blockchain-enabled cybersecurity and data integrity model, Scientific Reports 15 (2025) 23454.
\newblock \href {https://doi.org/10.1038/s41598-025-06405-y} {\path{doi:10.1038/s41598-025-06405-y}}.

\bibitem{nassar2020blockchain}
M.~Nassar, K.~Salah, M.~H. Ur~Rehman, D.~Svetinovic, Blockchain for explainable and trustworthy artificial intelligence, Wiley Interdisciplinary Reviews: Data Mining and Knowledge Discovery 10~(1) (2020) e1340.

\bibitem{LIU2024}
Blockchain and trusted reputation assessment-based incentive mechanism for healthcare services, Future Generation Computer Systems 154 (2024) 59--71.
\newblock \href {https://doi.org/https://doi.org/10.1016/j.future.2023.12.023} {\path{doi:https://doi.org/10.1016/j.future.2023.12.023}}.

\bibitem{SIDDIQUI2024}
S.~Siddiqui, S.~Hameed, S.~A. Shah, J.~Arshad, Y.~Ahmed, D.~Draheim, A smart-contract-based adaptive security governance architecture for smart city service interoperations, Sustainable Cities and Society 113 (2024) 105717.
\newblock \href {https://doi.org/https://doi.org/10.1016/j.scs.2024.105717} {\path{doi:https://doi.org/10.1016/j.scs.2024.105717}}.

\bibitem{mustafa2025blockchain}
G.~Mustafa, W.~Rafiq, N.~Jhamat, Z.~Arshad, F.~A. Rana, Blockchain-based governance models in e-government: a comprehensive framework for legal, technical, ethical and security considerations, International Journal of Law and Management 67~(1) (2025) 37--55.

\bibitem{SIDDIQUI2023}
S.~Siddiqui, S.~Hameed, S.~A. Shah, A.~K. Khan, A.~Aneiba, Smart contract-based security architecture for collaborative services in municipal smart cities, Journal of Systems Architecture 135 (2023) 102802.
\newblock \href {https://doi.org/https://doi.org/10.1016/j.sysarc.2022.102802} {\path{doi:https://doi.org/10.1016/j.sysarc.2022.102802}}.

\bibitem{konighofer2023}
B.~K{\"o}nighofer, J.~Rudolf, A.~Palmisano, M.~Tappler, R.~Bloem, Online shielding for reinforcement learning, Innovations in Systems and Software Engineering 19~(4) (2023) 379--394.
\newblock \href {https://doi.org/https://doi.org/10.1007/s11334-022-00480-4} {\path{doi:https://doi.org/10.1007/s11334-022-00480-4}}.

\bibitem{XACML12019}
T.~Kanwal, A.~A. Jabbar, A.~Anjum, S.~U. Malik, A.~Khan, N.~Ahmad, U.~Manzoor, M.~N. Shahzad, M.~A. Balubaid, Privacy-aware relationship semantics--based xacml access control model for electronic health records in hybrid cloud, International Journal of Distributed Sensor Networks 15~(6) (2019) 1550147719846050.

\bibitem{odedina2023redefining}
E.~A. Odedina, Redefining governance, risk, and compliance (grc) in the digital age: Integrating ai-driven risk management frameworks, World Journal of Advanced Engineering Technology and Sciences 10~(01) (2023) 264--282.
\newblock \href {https://doi.org/10.30574/wjaets.2023.10.1.0257} {\path{doi:10.30574/wjaets.2023.10.1.0257}}.

\bibitem{sadeghi2024interoperability}
M.~Sadeghi, A.~Carenini, O.~Corcho, M.~Rossi, R.~Santoro, A.~Vogelsang, et~al., Interoperability of heterogeneous systems of systems: from requirements to a reference architecture, The Journal of Supercomputing 80~(7) (2024) 8954--8987.

\bibitem{Alvin2020}
A.~Huseinović, S.~Mrdović, K.~Bicakci, S.~Uludag, A survey of denial-of-service attacks and solutions in the smart grid, IEEE Access 8 (2020) 177447--177470.
\newblock \href {https://doi.org/10.1109/ACCESS.2020.3026923} {\path{doi:10.1109/ACCESS.2020.3026923}}.

\bibitem{DEVITO2025}
G.~{De Vito}, F.~Palomba, F.~Ferrucci, The role of large language models in addressing iot challenges: A systematic literature review, Future Generation Computer Systems 171 (2025) 107829.
\newblock \href {https://doi.org/https://doi.org/10.1016/j.future.2025.107829} {\path{doi:https://doi.org/10.1016/j.future.2025.107829}}.

\bibitem{adewusi2024conceptual}
B.~A. Adewusi, B.~I. Adekunle, S.~D. Mustapha, A.~C. Uzoka, A conceptual model for responsible ai integration in public-facing digital services and platform governance, International Journal of Scientific Research in Computer Science, Engineering and Information Technology 10~(2) (2024) 416--435.
\newblock \href {https://doi.org/10.32628/CSEIT2425416735} {\path{doi:10.32628/CSEIT2425416735}}.

\bibitem{antuley2025securing}
U.~Antuley, S.~Hameed, S.~Siddiqui, S.~A. Shah, Securing smart city ecosystems: A taxonomy-based review of emerging technologies and frameworks for scalable collaborative services, IET Smart Cities 7~(1) (2025) e70007.
\newblock \href {https://doi.org/10.1049/smc2.70007} {\path{doi:10.1049/smc2.70007}}.

\bibitem{dazzi2025internet}
P.~Dazzi, The internet of ai agents (iaia): A new frontier in networked and distributed intelligence, International Journal of Networked and Distributed Computing 13 (2025) 16.
\newblock \href {https://doi.org/10.1007/s44227-025-00057-0} {\path{doi:10.1007/s44227-025-00057-0}}.

\bibitem{golpayegani2024adaptation}
F.~Golpayegani, N.~Chen, N.~Afraz, E.~Gyamfi, A.~Malekjafarian, D.~Sch{\"a}fer, C.~Krupitzer, Adaptation in edge computing: a review on design principles and research challenges, ACM Transactions on Autonomous and Adaptive Systems 19~(3) (2024) 1--43.

\bibitem{Ayub2024}
K.~Ayub, R.~alShawa, A secure iot framework for smart cities: Integrating servicenow irm/grc with blockchain and ai-driven threat detection, in: 2024 International Conference on Computer and Applications (ICCA), 2024, pp. 1--6.
\newblock \href {https://doi.org/10.1109/ICCA62237.2024.10927950} {\path{doi:10.1109/ICCA62237.2024.10927950}}.

\bibitem{Tong2020}
T.~Wu, P.~Zhou, K.~Liu, Y.~Yuan, X.~Wang, H.~Huang, D.~O. Wu, Multi-agent deep reinforcement learning for urban traffic light control in vehicular networks, IEEE Transactions on Vehicular Technology 69~(8) (2020) 8243--8256.
\newblock \href {https://doi.org/10.1109/TVT.2020.2997896} {\path{doi:10.1109/TVT.2020.2997896}}.

\bibitem{LI2021}
Z.~Li, H.~Yu, G.~Zhang, S.~Dong, C.-Z. Xu, Network-wide traffic signal control optimization using a multi-agent deep reinforcement learning, Transportation Research Part C: Emerging Technologies 125 (2021) 103059.
\newblock \href {https://doi.org/https://doi.org/10.1016/j.trc.2021.103059} {\path{doi:https://doi.org/10.1016/j.trc.2021.103059}}.

\bibitem{cai2024traffic}
C.~Cai, M.~Wei, Adaptive urban traffic signal control based on enhanced deep reinforcement learning, Scientific Reports 14 (2024) 14116.
\newblock \href {https://doi.org/10.1038/s41598-024-64885-w} {\path{doi:10.1038/s41598-024-64885-w}}.

\bibitem{hameed2023}
S.~Hameed, A.~Islam, K.~Ahmad, et~al., Deep learning based multimodal urban air quality prediction and traffic analytics, Scientific Reports 13 (2023) 22181.
\newblock \href {https://doi.org/10.1038/s41598-023-49296-7} {\path{doi:10.1038/s41598-023-49296-7}}.

\bibitem{Ozsoyeller2024}
D.~Ozsoyeller, Öznur Özkasap, Distributed asynchronous rendezvous planning on the line for multi-agent systems, Future Generation Computer Systems 161 (2024) 35--48.
\newblock \href {https://doi.org/https://doi.org/10.1016/j.future.2024.06.054} {\path{doi:https://doi.org/10.1016/j.future.2024.06.054}}.

\bibitem{LEE2025int}
J.~Lee, J.~Kim, S.~K. Yoo, T.~Taleb, J.~Song, Standardised interworking and deployment of iot and edge computing platforms, Digital Communications and Networks (2025).
\newblock \href {https://doi.org/https://doi.org/10.1016/j.dcan.2025.04.006} {\path{doi:https://doi.org/10.1016/j.dcan.2025.04.006}}.

\bibitem{ELAMANOV2024mod}
S.~Elamanov, H.~Son, B.~Flynn, S.~K. Yoo, N.~Dilshad, J.~Song, Interworking between modbus and internet of things platform for industrial services, Digital Communications and Networks 10~(2) (2024) 461--471.
\newblock \href {https://doi.org/https://doi.org/10.1016/j.dcan.2022.09.013} {\path{doi:https://doi.org/10.1016/j.dcan.2022.09.013}}.

\bibitem{Koenighofer2023}
B.~K{\"o}nighofer, J.~Rudolf, A.~Palmisano, M.~Alshiekh, E.~Bartocci, R.~Bloem, K.~Chatterjee, T.~A. Henzinger, A.~Turrini, Online shielding for reinforcement learning, Innovations in Systems and Software Engineering 19 (2023) 379--394.
\newblock \href {https://doi.org/10.1007/s11334-022-00480-4} {\path{doi:10.1007/s11334-022-00480-4}}.

\bibitem{Ali2025}
A.~Al-Haj, Enhancing iot security schemes through governance, risk, and compliance (grc), in: S.~Li (Ed.), Information Management, Springer Nature Switzerland, 2026, pp. 217--226.
\newblock \href {https://doi.org/10.1007/978-3-031-99353-4_19} {\path{doi:10.1007/978-3-031-99353-4_19}}.

\bibitem{sohail2025udt}
A.~Sohail, B.~Shen, M.~A. Cheema, et~al., Beyond data, towards sustainability: a sydney case study on urban digital twins, PFG 93 (2025) 365--377.
\newblock \href {https://doi.org/10.1007/s41064-025-00337-y} {\path{doi:10.1007/s41064-025-00337-y}}.

\bibitem{WANG2021}
T.~Wang, J.~Cao, A.~Hussain, Adaptive traffic signal control for large-scale scenario with cooperative group-based multi-agent reinforcement learning, Transportation Research Part C: Emerging Technologies 125 (2021) 103046.
\newblock \href {https://doi.org/https://doi.org/10.1016/j.trc.2021.103046} {\path{doi:https://doi.org/10.1016/j.trc.2021.103046}}.

\bibitem{haseeb2025comprehensive}
F.~Haseeb, S.~Ali, N.~Ahmed, N.~Alarifi, Y.~M. Youssef, Comprehensive probabilistic analysis and practical implications of rainfall distribution in pakistan, Atmosphere 16~(2) (2025) 122.
\newblock \href {https://doi.org/https://doi.org/10.3390/atmos16020122} {\path{doi:https://doi.org/10.3390/atmos16020122}}.

\bibitem{amjad2022analysis}
M.~Amjad, A.~Khan, K.~Fatima, O.~Ajaz, S.~Ali, K.~Main, Analysis of temperature variability, trends and prediction in the karachi region of pakistan using arima models, Atmosphere 14~(1) (2022) 88.

\bibitem{YOLOv8_Traffic}
F.~Nekouee, \href{https://github.com/FarzadNekouee/YOLOv8_Traffic_Density_Estimation}{Yolov8 traffic density estimation} (2023).
\newline\urlprefix\url{https://github.com/FarzadNekouee/YOLOv8_Traffic_Density_Estimation}

\bibitem{kausar2024evaluating}
A.~Kausar, S.~Zubair, H.~Sohail, et~al., Evaluating the challenges and impacts of mixed-use neighborhoods on urban planning: an empirical study of a megacity, karachi, pakistan, Discover Sustainability 5~(24) (2024) 1--14.
\newblock \href {https://doi.org/10.1007/s43621-024-00195-5} {\path{doi:10.1007/s43621-024-00195-5}}.

\bibitem{YOLO11_FireSmoke}
S.~Gamal, \href{https://github.com/sayedgamal99/Real-Time-Smoke-Fire-Detection-YOLO11}{Real-time smoke and fire detection with yolo11} (2024).
\newline\urlprefix\url{https://github.com/sayedgamal99/Real-Time-Smoke-Fire-Detection-YOLO11}

\bibitem{nist2023airmf}
{National Institute of Standards and Technology (NIST)}, \href{https://www.nist.gov/itl/ai-risk-management-framework}{Artificial intelligence risk management framework (ai rmf 1.0)}, functions: Govern, Map, Measure, Manage; emphasizes roles, responsibilities, and accountability (2023).
\newline\urlprefix\url{https://www.nist.gov/itl/ai-risk-management-framework}

\bibitem{chaudhary2025ai}
H.~Chaudhary, G.~Sharma, D.~Nishad, S.~Khalid, Ai-enhanced modelling of queueing and scheduling systems in cloud computing, Discover Applied Sciences 7~(4) (2025) 276.

\end{thebibliography}
%\input{elsarticleref.bbl}


\end{document}